\documentclass[11pt,english]{article}
\usepackage{mathptmx}
\usepackage{berasans}
\usepackage{beramono}

\usepackage[T1]{fontenc}
\usepackage[latin9]{inputenc}
\usepackage{geometry}
\usepackage{color}
\usepackage{babel}
\usepackage{float}
\usepackage{amsthm}
\usepackage{amsmath}
\usepackage{amssymb}
\usepackage{enumerate}
\usepackage{graphicx}
\usepackage{esint}
\usepackage{subfigure}
\usepackage{soul}
\usepackage[authoryear]{natbib}
\usepackage[unicode=true,
bookmarks=true,bookmarksnumbered=true,bookmarksopen=true,bookmarksopenlevel=2,
breaklinks=true,pdfborder={0 0 0},backref=false,colorlinks=true]
{hyperref}
\hypersetup{pdftitle={The Asymptotic Additive Properties of Value-at-Risk under Copula Tail Dependence},
	pdfauthor={Wenhao Zhu},
	pdfsubject={2017},
	pdfkeywords={ Value-at-Risk; asymptotic diversification ratio; asymptotic sub-additivity; tail concave order; tail dependence function},
	linkcolor=blue,anchorcolor=blue,citecolor=blue}

\makeatletter

\theoremstyle{definition}

\theoremstyle{definition}
\newtheorem{defn}{\protect\definitionname}[section]
\theoremstyle{plain}
\newtheorem{prop}{\protect\propositionname}[section]
\theoremstyle{definition}
\newtheorem{example}{\protect\examplename}[section]
\ifx\proof\undefined\
\newenvironment{proof}[1][\proofname]{\par
	\normalfont\topsep6\p@\@plus6\p@\relax
	\trivlist
	\itemindent\parindent
	\item[\hskip\labelsep
	\scshape
	#1]\ignorespaces
}{%
	\endtrivlist\@endpefalse
}
\providecommand{\proofname}{Proof}
\fi
\theoremstyle{plain}

\theoremstyle{remark}
\newtheorem{rem}{\protect\remarkname}[section]
\theoremstyle{plain}
\newtheorem{thm}{\protect\theoremname}[section]
\theoremstyle{plain}
\newtheorem{cor}{\protect\corollaryname}[section]
\theoremstyle{plain}

\usepackage{indentfirst}

\makeatother
\providecommand{\assumptionname}{Assumption}
\providecommand{\corollaryname}{Corollary}
\providecommand{\definitionname}{Definition}
\providecommand{\examplename}{Example}
\providecommand{\lemmaname}{Lemma}
\providecommand{\propositionname}{Proposition}
\providecommand{\remarkname}{Remark}
\providecommand{\theoremname}{Theorem}
\providecommand{\conditionname}{Condition}

\makeatletter

\newcommand{\Rmnum}[1]{\expandafter\@slowromancap\romannumeral #1@}
\makeatother

\begin{document}
	
	\title{\textbf{Risk aversion with two risks  for  state-dependent  utilities}}
	
	\author{\textit{Jiehua Xie$^{\rm a}$,\ \ Biyao Wu$^{\rm b}$,\ \ Wei Zou$^{\rm c}$\thanks{Corresponding
				author. \textit{Email-address:}zouwei@juwp.edu.cn (Wei Zou).} }
		\\
\small \textit{$^{\rm a}$School of Business Administration, Jiangxi University of Water Resources and Electric Power, Jiangxi 330099, P.R. China}\\	\small \textit{$^{\rm b}$School of Information Management, Wuhan University, Hubei 430072, P.R.China}	
\\
\small \textit{$^{\rm c}$School of Science, Jiangxi University of Water Resources and Electric Power, Jiangxi 330099, P.R. China}
}
		\date{August, 2026}
	\maketitle
	
	\begin{abstract}

In this paper, we introduce the concept of Arrow-Pratt risk aversion within a two-attribute, state-dependent utility framework, where a primary risk is accompanied by a background risk. Within this setting, we investigate the interrelationships among risk aversion toward one risk in the presence of another risk, the structure of state-dependent two-attribute utility functions, and the degree of correlation between the two risks. More specifically, we establish conditions on the state-dependent two-attribute utility function under which risk aversion toward primary risk is preserved when a background risk is involved, considering various correlation structures between the two types of risks. Additionally, a new concept, uniform risk aversion in the presence of a background risk, is introduced. We show that the above established conditions on the state-dependent two-attribute  utility function are equivalent to uniform risk aversion toward primary risk in presence of a background risk. These findings offer new insights into state-dependent preferences for correlated risks. Two applications in the contexts of fair insurance contract and optimal prevention within a two-period model  illustrate the practical usefulness of the proposed  risk aversion with two risks under state-dependent utilities.
	\end{abstract}
	
	\section{Introduction}
Risk aversion has long been one of fundamental concepts in  decision-making under risk and uncertainty. In their seminal works,  
\cite{Essays1971} and \cite{1964Pratt} formally defined the risk aversion as the preference for the certain receipt of a risk's expected value over the risk itself, within the framework
of expected utility. Under this formulation, the degree of risk aversion among different decision makers (DMs) can be systematically  compared through their respective utility functions. Subsequent research has considerably broadened and deepened the theoretical framework of risk aversion \citep{1969Some,1974Increases,1981Some}.  


In the classical theory of risk aversion, it is typically assumed that the utility 
of outcomes is \textit{state-independent}, implying that a DM's utility does  not vary with the state of nature. This
assumption substantially simplifies the identification and comparison of utility functions across heterogeneous decision environments. However, as noted  by \cite{1983On}, there exist situations in which ``the evaluation of the consequences is not independent of the
prevailing state of nature''. For example, an individual who has sustained a permanent injury may assign a different utility to a monetary endowment of $\$$1 million than when in good health. Such situations illustrate that DMs may exhibit  \textit{state-dependent} utilities. The literature on risk aversion for state-dependent utilities is abundant \citep{1983aversion,1983On,1991Risk,1991A,2003A,2021Jarrow,2025Belief}. To highlight the importance and broad applicability of this topic, we start with several examples taken from the existing literature.


In the context of insurance, the assumption of  state-independent utility is often inconsistent with some practical settings, such as  life, health, or disability insurance. An important implication of  this inconsistency  is that a risk averse individual may reject full insurance coverage even when it is offered at actuarially fair terms \citep{1977The}. Consequently, it has been argued that a state-dependent utility framework provides a more realistic and accurate representation of decision-making behavior in these types of insurance contexts \citep{1983aversion,1996Risk}.

In the field of behavioral economics, numerous examples also illustrate that risk attitudes of DMs often depend on the  circumstances in which decisions are made. Empirical and experimental evidence from \cite{1991A}, \cite{2003A}, and
\cite{2015Testing}, among others, indicates that identical outcomes may be evaluated differently depending on the environmental states  in which they occur. Hence, the assumption of state-independent risk attitudes
may result in inaccurate assessments of risk and lead to suboptimal
decision outcomes.

In financial economics, as emphasized by \cite{2021Jarrow}, the use of state-dependent utility functions  is important  for the following reasons: (i) it enables the 
consumption and portfolio choice theory to be consistent with some of the observed behavioral violations of optimal and rational  investor behavior under state-independent preferences \citep{Zitkovic2012Utility,2017Optimal,2017Mo}; (ii) in an incomplete market, the state-dependent utility functions are essential to characterize the economic equilibrium \citep{2000A,2012Rational,2017On}. 


It is noteworthy that these studies discussed above, along with most of the existing research on risk aversion under state-dependent utilities, have predominantly been conducted within the framework of single-attribute utility functions. In practice, however, DMs often face multiple sources of risk simultaneously. Among these, a primary or endogenous risk, such as financial or insurable risk, typically coexists with exogenous sources of uncertainty, including macroeconomic, health, or environmental risks. These exogenous factors are frequently combined together and treated as a \textit{background risk} \citep{1988Aversion,2006Multiplicative,2011Risk}. 
In many real-world settings, the two risks originate from different  dimensions, thereby motivating the adoption of two-attribute utility functions to analyze how background risk influences the preservation of risk aversion toward a primarily risk  \citep{1999On,2011Correlated,2016Preserving,2017Preserving,2016Risk,2021Comparative}. Despite these notable results, the research has not systematically incorporate the state-dependent bivariate (or two-attribute) utility function (SDBUF) into the analysis of DMs' attitude toward a primary risk in the presence of a background risk. This paper aims to fill this gap by extending the theoretical analysis of Arrow-Pratt risk aversion under joint risk exposure through the development and application of the SDBUF framework.

To illustrate the central themes of this study, we consider a DM  in good health who makes current prevention activities, such as saving, regular medical checkups, or lifestyle improvements, to enhance both future wealth and health status. In this context, wealth and health status represent two distinct dimensions of well-being, and the two variables are generally correlated. At the time of making this decision, the individual knows his/her current wealth, health status, and utility level. However, uncertainty remains regarding the future   realizations of the three variables.  Within this setting, we are interested to investigate how the correlation between the future values of these two attributes (wealth and health status), together with the state-dependent utilities, influences the DMs' preferences toward correlated risks and, consequently, their prevention activities. 

Building upon these insights, we study the risk aversion toward one primary risk, such as a financial or insurable risk, in the presence of a background risk  under the SDBUF framework. Here, a DM's preference is characterized by his/her expected SDBUF, and the DM 
is said to be risk averse toward a primary risk $X$ in the presence of a background risk $Y$ if he/she prefers $(\mathbb{E}_{\mathbb{P}}[X],Y)$ to $(X,Y)$.  The relations between the risk aversion with two risks, the structure of SDBUF and  the degree of correlation between the two  risks are discussed. Through these investigations, we offer some new theoretical insights into state-dependent preferences toward two  correlated risks. The results developed in this paper extend existing research that focused on single-risk and mainly used state-dependent single-attribute utility functions \citep[see, e.g.,][]{2021Jarrow}. The main works of this paper are twofold.


First, we discuss the case in which the considered primary and background risks are statistically independent\footnote{To avoid the confusion with the concept of utility independence in mulattribute utility theory, the independence between random variables is termed statistical independence in this paper.} of the SDBUF's state space. When no constraint is imposed on the correlation between the two risks, it is shown that the additive form of the SDBUF, corresponding to the statewise \textit{mutual utility independence} \citep{Keeney1976}, implies the risk aversion to one risk in the
presence of another. When the two risks are statistically  independent,  we verify that the statewise concavity of SDBUF  toward the primary risk alone guarantees the risk aversion toward the same risk in the
presence of a background risk. In the case where the primary risk exhibits \textit{positive expectation-dependence} \citep{1987Expectation,2011The} on the background risk, we establish the orders of \textit{risk apportionment} \citep{2007A,2011J} on the SDBUF required  to preserve risk aversion with two risks. A similar result is also presented when the primary risk is \textit{positively regression-dependent} \citep{2005Risky} on the background risk. To deepen this analysis, we introduce a new concept of bivariate risk aversion, termed \textit{uniform risk aversion in  presence of a background risk}. We verify that the above  established conditions on the SDBUF are equivalent to the notation of uniform risk aversion in the presence of a background risk under the corresponding correlation structures  between the two risks. This equivalence offers additional economic interpretation and behavioral insight into  these established conditions on the SDBUF.

Second, we  address the case in which the state space of the SDBUF  coincides with that of risks. We show that the conditions on SDBUF established in the above case are not applicable in this state-dependent setting. A sufficient condition guaranteeing the preservation of risk aversion toward the primary risk in the presence of a background risk is established within this state-dependent utility framework. 
The economic interpretation of this condition is also clarified. Moreover, when the SDBUF has a \textit{utility dominant attribute}  \citep{2005Attribute}, we present the equivalent conditions for uniform risk aversion in the presence of a background risk under various forms of correlation  between the two risks in this state-dependent case.  


In summary, this study combines four concepts: ($i$) risk aversion to one risk in the presence of another, ($ii$) risk
apportionment as formulated by \cite{2006Putting}, ($iii$) utility independence \citep{Keeney1976} and attribute dominance utility \citep{2005Attribute}, 
and ($iv$) SDBUFs. Consequently, the main contribution of this paper lies
not only in extending existing analyses but also in establishing a unified theoretical framework for various concepts.

The remainder of the paper is organized as follows. In Section \ref{sec:Tail Dependence function}, we present the basic notations and
preliminaries, including  various forms of correlation  between two risks, the notation of risk apportionment, and key properties of utility functions. In Section \ref{chaper:risk aversion}, the case in which the  risks are statistically independent of the SDBUF's state space is discussed. We establish the conditions on SDBUFs implying risk aversion toward one risk in the presence of a background risk  under different correlation structures between the two risks. We also verify  that these established conditions on SDBUFs are equivalent to the concept of uniform bivariate risk aversion. Section \ref{chaper:uniform} addresses the case in which the state space of SDBUF coincides with that of risks. A sufficient condition guaranteeing the preservation of risk aversion to a primarily risk in the presence of a background risk for this state-dependent case is presented, and its economic interpretation is subsequently examined. Furthermore, the equivalent conditions for uniform bivariate risk aversion in this state-dependent framework are also identified. In Section \ref{applications}, 
two applications in the contexts of fair insurance contracts and  optimal prevention within a two-period model are provided to illustrate the practical usefulness of the proposed concepts and conditions of bivariate risk aversion for state-dependent utilities.  The
conclusions are drawn in Section \ref{sec:Conclusion}.  All  proofs are put in Appendix.

\section{Notations and  preliminaries}\label{sec:Tail Dependence function}

\subsection{The correlation between two risks}

Since an individual  may face multiple sources of risks simultaneously, then two different types of risk,  the primary risk and the background risk are considered. Let $X$ be the primary risk, such as  an insurable risk or a financial risk, and $Y$  be the background risk. We consider a probability space $(\Omega, \mathcal{F}, \mathbb{P} )$ representing the stochastic environment in which risks are defined, where  $\mathbb{P}$ denotes the DM's subjective probability measure toward risks. 

For risks $X$ and $Y$, the cumulative distribution function (CDF) of random vector $(X,Y)$ over support $ [\underline{x},\overline{x}]\times[\underline{y},\overline{y}]$ is denoted by $F^{\mathbb{P}}(x,y)$, where $\underline{x}<\overline{x}$ and $\underline{y}<\overline{y}$. Here, $\underline{x}$ (or $\underline{y}$) may be a finite constant or $-\infty$, and $\overline{x}$ (or $\overline{y}$) may be a finite constant or $+\infty$. The marginal distributions of $X$ and $Y$ over their respective supports $[\underline{x},\overline{x}]$ and $[\underline{y},\overline{y}]$ are denoted by  $F_X^{\mathbb{P}}(x)$ and $F_Y^{\mathbb{P}}(y)$, respectively. For any probability measure $Q$, let $\mathbb{E}_{Q}[\cdot]$ be the expectation under  $Q$ and $\mathbb{E}_{Q}[\cdot|A]$ be the conditional expectation given the occurrence of event $A$. 

The correlation between risks $X$ and $Y$ plays an important role in the preservation of risk aversion toward one risk in the presence of another risk \citep{1999On}.  In the following, for risks $X$ and $Y$, several different forms of correlation are introduced.

\begin{defn}\label{de:appoint-1} \citep[ Definition 1]{2005Risky} The risk $X$ is positively regression-dependent  on $Y$, denoted as $(X,Y)\in R^{\mathbb{P}}(X|Y)$, if $\mathbb{E}_{\mathbb{P}}[X|Y=y]$ is non-decreasing with respect to $y$ on  $[\underline{y},\overline{y}]$.
\end{defn}
The positive regression-dependence (PRD, in short) introduced in  Definition \ref{de:appoint-1} is a widespread type of correlation  between two risks \citep{2005Risky}. For example, when $X$ and $Y$ are statistically independent risks or comonotonic risks, $(X,Y)\in R^{\mathbb{P}}(X|Y)$.

Denote 
$$\mathbb{E}_{\mathbb{P}}D_1(X|y):=\mathbb{E}_{\mathbb{P}}[X]-\mathbb{E}_{\mathbb{P}}\big[X|Y\le y\big]\ \ {\rm for\ all}\ \ y\in[\underline{y}, \overline{y}].$$
\cite{1987Expectation} introduced the following form of correlation between two risks, which is also widely used in the literature.

\begin{defn}\label{de:appoint-2}  The risk $X$ is positively  first-degree expectation-dependent ($1$st-degree PED, in short) on $Y$, denoted as $(X,Y)\in \mathbb{E}_{\mathbb{P}}D_{1}(X|Y)$, if  $\mathbb{E}_{\mathbb{P}}D_{1}(X|y)\ge 0$ for all $y\in[\underline{y}, \overline{y}]$. 
\end{defn} 
\cite{1987Expectation} verified that $1$st-degree PED is a weaker correlation than PRD. Consequently, 
\begin{equation}\label{dependence1}
(X,Y)\in R^{\mathbb{P}}(X|Y)\Rightarrow (X,Y)\in \mathbb{E}_{\mathbb{P}}D_{1}(X|Y).
\end{equation} 

\cite{2011The} extended \cite{1987Expectation}'s notion of $1$st-degree PED
to second-degree by using the following function:
$$\mathbb{E}_{\mathbb{P}}D_2(X|y)=\int_{\underline{y}}^y \mathbb{E}_{\mathbb{P}}D_1(X|t)F^{\mathbb{P}}_Y(t){\rm d}t.$$
For $N\ge 3$, \cite{2011The} also defined the repeated integrals as 
$$\mathbb{E}_{\mathbb{P}}D_N(X|y)=\int_{\underline{y}}^y \mathbb{E}_{\mathbb{P}}D_{N-1}(X|t){\rm d}t,\ N\ge 3,$$
and then introduced the positive $N$th-degree expectation-dependence ($N$th-degree PED, in short). 
\begin{defn}\label{de:appoint}  For $N>1$, $X$ is $N$th-degree PED on $Y$,  denoted as $(X,Y)\in \mathbb{E}_{\mathbb{P}}D_{N}(X|Y)$, if $\mathbb{E}_{\mathbb{P}}D_{k+1}(X|\overline y)\ge 0$ for each $k=1,\ldots,N-1$ and $\mathbb{E}_{\mathbb{P}}D_{N}(X| y)\ge 0$ for all $y\in[\underline y, \overline y]$.	
\end{defn}

\cite{2011The} showed that  
\begin{equation}\label{dependence2} (X,Y)\in \mathbb{E}_{\mathbb{P}}D_{1}(X|Y)\Rightarrow (X,Y)\in \mathbb{E}_{\mathbb{P}}D_{2}(X |Y)\Rightarrow\cdots\Rightarrow (X,Y)\in \mathbb{E}_{\mathbb{P}}D_{N}(X|Y),\ N> 1.\end{equation}
Thus,  for the forms of correlation between two risks mentioned above,  $N$th-degree PED is
the weakest and PRD is the strongest.

\subsection{Utility function}\label{Utility function}

Since both primary risk $X$ and background risk $Y$ are considered, the utility function $U$ of a DM is assumed to be two-attribute, that is, $U(x,y)$, $(x,y)\in[\underline{x},\overline{x}]\times[\underline{y},\overline{y}]$ is a two-attribute (or bivariate) utility function. 

In decisions with multiple attributes, the utility independence is one fundamental concept for decomposing the multiattribute utility function
into single-attribute utility functions \citep{Keeney1976}. In particular, for a two-attribute utility function $U(x,y)$, if each attribute is utility independent of its complement (i.e., mutual utility independence), then $U(x,y)$ can be represented in either an additive form
\begin{equation}
	U(x,y)=k_1U_1(x)+k_2U_2(y) 
\end{equation} 
or a multiplicative form $1+kU(x,y)=\big(1+kk_1U_1(x)\big)\big(1+kk_2U_2(y)\big)$, where $k_1$, $k_2$ are scaling constants, and $U_i$, $i=1,2$ are single-attribute utility functions.  All these
forms of two-attribute utility functions are widely
employed in both theoretical and applied analyses owing to their analytical tractability and practical simplicity  \citep{Basak2014Strategic,xie2022patchwork}.

Next, another class of two-attribute utility functions is introduced.
\begin{defn}\label{de:Dominance} \citep[]{2005Attribute} A two-attribute utility function $U(x,y)$ is said to be an attribute dominance
	utility function if 
	$U(\underline{x},y)=U(x,\underline{y})=U(\underline{x},\underline{y})\  {\rm for\ all}\ (x,y)\in[\underline{x},\overline{x}]\times[\underline{y},\overline{y}]$. An attribute $x$ is referred to as a utility dominant attribute if
	$U(\underline{x},y)=U(\underline{x},\underline{y})\ \ {\rm for\ all}\ y\in[\underline{y},\overline{y}]$. 
\end{defn}

Attribute dominance utility functions frequently appear in  decision-making applications, particularly in contexts involving  life and death situations, where these attributes may represent  consumption level and health state \citep{On1980}.

Throughout this paper,  we use the notation $U^{(n,m)}(x, y):=\frac{\partial^{n+m} U(x, y)}{\partial x^n \partial y^m}$   
to denote the $(m + n)$th-order partial derivative
of $U(x, y)$ with respect to $x$ and $y$, 
where $n$ and $m$ are non-negative integers. Following the definition introduced in \cite{2006Putting}, \cite{2007A} and \cite{2011J}, we briefly recall the concept of risk apportionment of order $(n, m)$, which is employed to explain the signs of $U^{(n,m)}$. Let $\tilde{\epsilon}_i$, $i=1,2,\ldots$, be statistically independent random variables that are also statistically independent of DM's wealth. Define the base cases $A_1=[-k]$ and  $B_1=B_2=[0]$,  and let $A_2=[\tilde{\epsilon}_1]$.  \cite{2006Putting} generalized these definitions to higher orders through the following recursive construction of 50-50 lotteries:
$$A_n=[0+B_{n-2};\tilde{\epsilon}_{{\rm Int}(n/2)}+A_{n-2}]\ \ \ {\rm and}\ \ \ B_n=[0+A_{n-2};\tilde{\epsilon}_{{\rm Int}(n/2)}+B_{n-2}] $$
for each $n\ge 3$. Furthermore, \cite{2011J} extended this framework by proposing a definition of bivariate risk apportionment.

\begin{defn}\label{de:risk}  Preferences exhibit risk apportionment of order $(M,N)$, if for all $x,y$, the DM prefers the lottery $[(x+B_{N}, y+A_{M});(x+A_{N}, y+B_{M})]$ to $[(x+B_{N}, y+B_{M});(x+A_{N}, y+A_{M})]$.
\end{defn}

Intuitively, when an individual tends to disaggregate risks, his/her
preference exhibits risk apportionment. \cite{2011J} verified that
bivariate risk apportionment can be characterized in terms of the cross-partial derivatives
of DM's utility function $U$.

\begin{thm} \citep[]{2011J}\label{thmrisk}
	A two-attribute utility function $U$
	obeys risk apportionment of order $(M, N)$ if and only if, for all $(x, y)$, $(-1)^{M+N}U^{(M,N)}(x,y)\le 0$.
\end{thm}

Now, we consider a probability space $(\tilde{\Omega}, \tilde{\mathcal{F}}, \tilde{\mathbb{P}} )$ and assume that DM's utility function $U$ depends explicitly on the state space $\tilde{\Omega}$. That is, the DM holds a subjective belief $\tilde{\mathbb{P}}$ over a SDBUF $U(x,y,\tilde{\omega})$. 
For a more detailed introduction about state-dependent  utility function and its application, we refer to  \cite{Zitkovic2012Utility}, \cite{2017Optimal}, \cite{2017Mo} and \cite{2021Jarrow}. Throughout this paper, 
we consider the risks $X$, $Y$ and the SDBUF $U$, satisfying the conditions that $\mathbb{E}_{\mathbb{P}}\big[X\big]<\infty$ and  $\mathbb{E}_{\mathbb{P}\times \tilde{\mathbb{P}}}\big[|U\big(X, Y,\tilde{\omega}\big)|\big]<\infty$. For clarity, the properties of the utility function $U(x,y,\tilde{\Omega})$  evaluated at a given state $\tilde{\omega}\in \tilde{\Omega}$ are referred to as \textit{statewise} characteristics. The term ``statewise'' thus emphasizes the analysis conditioned on a fixed  state $\tilde{\omega}\in \tilde{\Omega}$. In the subsequent analysis, we focus on the following two cases within the SDBUF framework. 
\begin{itemize}
	\item[Case $I$.] The state spaces $\Omega$ and $\tilde{\Omega}$ are statistically independent. In this setting,  
	the realizations of risks $X$ and $Y$ are not  influenced by the state-dependence of the utility function on $\tilde{\Omega}$. The constructions of such SDBUFs and the associated choice set of risks are motivated  by applications of financial economics, particularly in asset pricing theory. In this context, $\omega\in\Omega$ typically  represents uncertainty relevant to the consumption or  terminal wealth, while $\tilde{\omega}\in\tilde{\Omega}$ captures an additional, statistically independent source of randomness. Incorporating this statistically independent source of randomness facilitates the study of the statewise characteristic of the SDBUF $U$. Within the literature on state-dependent preferences, this state-dependent framework has been employed primarily for single-attribute utility functions \citep[see, e.g.,][]{2017Optimal,2017Mo,2019Bayesian,2021Jarrow}.  
	
\item[Case $II$.] $(\tilde{\Omega}, \tilde{\mathcal{F}}, \tilde{\mathbb{P}} )=(\Omega, \mathcal{F}, \mathbb{P} )$, i.e., the probability spaces $\Omega$ and  $\tilde{\Omega}$ coincide. In this case, both the random variables $X$ as well as $Y$ and the utility function $U$ depend on the same state space $\Omega$. Consequently, the realizations of risks $X$ and $Y$ are influenced by the  state-dependence of the utility function $U$ on $\Omega$. This framework has also  been extensively applied in the literature on state-dependent preferences under single-attribute settings \citep[see, e.g.,][]{1983aversion,1983On,2019A,2021Jarrow,2025Belief}.

\end{itemize}

 \section{Bivariate risk aversion for independent lotteries}\label{chaper:risk aversion}

In this section, we focus on  Case $I$, where the state spaces $\Omega$ and $\tilde{\Omega}$ are assumed to be statistically independent. Within this framework, we establish the conditions under which  risk
aversion toward a primary risk $X$ is preserved in the presence of a background risk  $Y$, when preferences are characterized by a  SDBUF $U(x,y,\tilde{\omega})$.

\subsection{Bivariate risk aversion for state-dependent utilities $U(x,y,\tilde{\omega})$  }\label{3.1}

We begin with a definition of aversion to one risk in presence of a background risk for a SDBUF $U(x,y,\tilde{\omega})$.

\begin{defn}\label{risk-aversion} 
	A SDBUF   $U(x,y,\tilde{\omega})$ is said to be risk averse to  $X$ in the presence of a background risk $Y$ if the inequality \begin{equation}\label{eq:1++}\mathbb{E}_{\mathbb{P}\times \tilde{\mathbb{P}}}\big[U\big(X, Y,\tilde{\omega}\big)\big]\le  \mathbb{E}_{\mathbb{P}\times\tilde{\mathbb{P}}}\big[U\big(\mathbb{E}_{\mathbb{P}}[X], Y,\tilde{\omega}\big)\big]\end{equation}
	holds. 
\end{defn}

The inequality \eqref{eq:1++} implies that, when the lotteries $X$ and $Y$ are  statistically  independent of the SDBUF $U(x, y, \tilde{\omega})$,  an agent with the utility $U$ prefers  $(\mathbb{E}_{\mathbb{P}}[X],Y)$
to the risks $(X,Y)$. This preference reflects  risk aversion toward the primary risk $X$ in the presence of a background risk $Y$ under state-dependent utilities. 
For brevity, the SDBUF $U(x,y,\tilde{\omega})$ is also said to be bivariate risk averse toward $(X,Y)$ whenever  the inequality \eqref{eq:1++} holds.

In what follows, we examine the conditions on the SDBUF  $U(x,y,\tilde{\omega})$ under which the  inequality \eqref{eq:1++} is satisfied for different correlation structures between $X$ and $Y$. We begin by analyzing the case where no constraints are imposed on the correlation between $X$ and $Y$. 

\begin{prop}\label{cor:Uindependent-gamble-sufficient}
For an arbitrary pair of risks $(X,Y)$, if for each state  $\tilde{\omega}\in\tilde{\Omega}$ and for all $(x,y)\in [\underline{x},\overline{x}]\times[\underline{y},\overline{y}]$, 
	$U(x,y,\tilde{\omega})=f(x,\tilde{\omega})+g(y,\tilde{\omega})$, where $f(x,\tilde{\omega})$ is statewise concave in $x$,  then $U$ is risk averse to $X$ in the presence of $Y$.
\end{prop}

As discussed in Subsection \ref{Utility function}, the additive form   $U(x,y,\tilde{\omega})=f(x,\tilde{\omega})+g(y,\tilde{\omega})$ implies the statewise utility independence between the two attributes $x$ and $y$. Consequently,  Proposition \ref{cor:Uindependent-gamble-sufficient} establishes that, in order to guarantee bivariate risk aversion toward arbitrary pair of risks  $(X,Y)$, the SDBUF $U(x,y,\tilde{\omega})$ must exhibit statewise utility independence, i.e., utility independence holds for every state $\tilde{\omega}\in \tilde{\Omega}$.  

The conditions required on the SDBUF in Proposition \ref{cor:Uindependent-gamble-sufficient} are relatively restrictive, as they are applicable to all pairs of risks $(X,Y)$, thereby imposing no constraint on the correlation between $X$ and $Y$.  Recall that  PRD  and $N$th-degree PED are widespread forms of correlation in the analysis and comparison of risks in the presence of a  background risk \citep{1987Expectation,2005Risky,2011The,2016Preserving,2017Preserving,2021Comparative,zhu2023Asymptotic}. In the following, under these specific  correlation structures, we establish the conditions on the SDBUF  $U(x,y,\tilde{\omega})$ that ensure the validity of  the  inequality \eqref{eq:1++}.


\begin{prop}\label{cor:Uindependent-gamble-sufficient+}
($i$).	For any risks $(X,Y)\in R^{\mathbb{P}}(X|Y)$, if for each state  $\tilde{\omega}\in\tilde{\Omega}$,  $U(x,y,\tilde{\omega})$ is concave it its first argument $x$  and  $U(x,y,\tilde{\omega})$
obeys risk apportionment of order $(1,1)$,  then $U$ is risk averse to $X$ in the presence of $Y$.

($ii$).	For any risks $(X,Y)\in \mathbb{E}_{\mathbb{P}}D_{N}(X|Y)$, if for each state  $\tilde{\omega}\in\tilde{\Omega}$,  $U(x,y,\tilde{\omega})$ is concave in its first argument $x$  and  $U(x,y,\tilde{\omega})$
obeys risk apportionment of order $(1,k)$ for $k\in\{1,\ldots, N\}$,  then $U$ is risk averse to  $X$ in the presence of  $Y$.
\end{prop}

It follows from  
Proposition \ref{cor:Uindependent-gamble-sufficient+}($ii$) and Theorem \ref{thmrisk} that, when a risk $X$ is $N$th-degree PED on the background risk $Y$, the extent to which risk aversion toward $X$ is preserved in the presence of $Y$ depends, for each state $\tilde{\omega}\in\tilde{\Omega}$, on the signs of the first $N$ partial derivatives of $U^{(1,0)}(x,y,\tilde{\omega})$ 
with respect to $y$. These derivatives characterize the admissible correlation structures between $X$ and $Y$ that guarantee bivariate risk aversion toward $(X,Y)$.

It is noteworthy that in Proposition \ref{cor:Uindependent-gamble-sufficient+}($ii$), when $N = 1$, the bivariate risk aversion conditions for $(X,Y)\in \mathbb{E}_{\mathbb{P}}D_{1}(X|Y)$ coincide with these for $(X,Y)\in R^{\mathbb{P}}(X|Y)$. However, Proposition \ref{cor:Uindependent-gamble-sufficient+}($ii$) represents a generalization of Proposition \ref{cor:Uindependent-gamble-sufficient+}($i$), as it requires less restrictive assumptions on the correlation between the two risks  
(that is, $ (X,Y)\in R^{\mathbb{P}}(X|Y)\Rightarrow$$(X,Y)\in \mathbb{E}_{\mathbb{P}}D_{1}(X|Y)$).  

Furthermore, since bivariate Bernoulli and bivariate normal pairs belong to the class $R^{\mathbb{P}}(X|Y)$, 
Proposition \ref{cor:Uindependent-gamble-sufficient+}($i$) directly  provides conditions on the SDBUF $U(x,y,\tilde{\omega})$ under which the  inequality \eqref{eq:1++} holds for these two classes of distributions.

\begin{cor}\label{cor1}
	If for each state  $\tilde{\omega}\in\tilde{\Omega}$, $U(x,y,\tilde{\omega})$ is concave in its first argument $x$  and  $U(x,y,\tilde{\omega})$ obeys risk apportionment of order $(1,1)$,  then $U$ is risk averse to $X$ in the presence of a background risk $Y$ whenever the pair $(X,Y)$ follows one of the following joint distributions: ($i$) a bivariate Bernoulli distribution with a non-negative correlation coefficient; ($ii$) a bivariate normal distribution with a non-negative coefficient correlation. 
\end{cor}

Observe that when risks $X$ and  $Y$ are statistically independent, the conditional expectation $\mathbb{E}_{\mathbb{P}}[X|Y=y]$ is non-decreasing in $y$, implying that $(X,Y)\in R^{\mathbb{P}}(X|Y)$. Thus, it follows from Proposition \ref{cor:Uindependent-gamble-sufficient+}($i$) that if for each state  $\tilde{\omega}\in\tilde{\Omega}$, $U(x,y,\tilde{\omega})$ is concave in its first argument $x$  and $U(x,y,\tilde{\omega})$ obeys risk apportionment of order $(1,1)$, then $U$ exhibits bivariate risk aversion toward $(X,Y)$. In fact, when risks $X$ and  $Y$ are statistically independent, the additional requirement that  $U(x,y,\tilde{\omega})$ obeys risk apportionment of order $(1,1)$ is not necessary. 

\begin{prop}\label{cor:Uindependent-gamble-sufficient++}
	For statistically independent risks $X$ and $Y$, if for each state  $\tilde{\omega}\in\tilde{\Omega}$,  $U(x,y,\tilde{\omega})$ is concave in its first argument $x$,  then $U$ is risk averse to $X$ in the presence of $Y$. 
\end{prop}

Propositions \ref{cor:Uindependent-gamble-sufficient}-\ref{cor:Uindependent-gamble-sufficient++} highlight the  role of the correlation between $X$ and $Y$  in the preserving risk aversion to  $X$ in the presence of $Y$. These results  show that the conditions ensuring bivariate risk aversion toward $(X,Y)$ for the SDBUF $U(x,y,\tilde{\omega})$ vary substantially depending on the form of  the correlation between $X$ and $Y$. In particular, as the class of  correlation structures between $X$ and $Y$  is expanded, the corresponding class of SDBUFs satisfying  the inequality \eqref{eq:1++} becomes  more restrictive. This reflects a form of duality between the correlation structure of risks and the characteristics of the SDBUF $U$.  Specifically, when statewise utility independence is achieved by the SDBUF, the risks $(X,Y)$ may exhibit arbitrary correlation. Conversely, when the statistical independence between $X$ and $Y$ is assumed, the functional form of SDBUF $U$ can be flexibly specified, subject only to the requirement of statewise concavity with respect to its first argument. Proposition  \ref{cor:Uindependent-gamble-sufficient+} presents an intermediate case, in which independence is not assumed either for the risks or for the SDBUF. In this framework, however, both the correlation structure of risks and the properties of SDBUF  are subject to specific constraints.

It is also important to note  that the converse statements to Propositions \ref{cor:Uindependent-gamble-sufficient}-\ref{cor:Uindependent-gamble-sufficient++} do not necessarily hold. To clarify this point, we  take Proposition \ref{cor:Uindependent-gamble-sufficient} as an illustrative example.

\begin{example}

Let $\Omega=\{\omega_1,\omega_2\}$, $\tilde{\Omega}=\{\tilde{\omega}_1,\tilde{\omega}_2\}$ and a SDBUF   $U(x,y,\tilde{\omega})$  be 
\begin{equation}\label{eq:utility}
	U(x,y,\tilde{\omega})= (x^{1/2}+y) \mathbb{I}_{\{\tilde{\omega}=\tilde{\omega}_1\}}(\tilde{\omega})+(x^{2}+y)\mathbb{I}_{\{\tilde{\omega}=\tilde{\omega}_2\}}(\tilde{\omega}),
\end{equation}
where  $\mathbb{I}_{A}$ is an indicator function of the set $A$, that is,	$\mathbb{I}_{A}$ equals 1 when $A$ is true and equals 0 otherwise. Obviously, $U(x,y,\tilde{\omega})$ is not concave in $x$ for the state $\tilde{\omega}=\tilde{\omega}_2$.

Now, suppose that 
$\mathbb{P}\times \tilde{\mathbb{P}}(\omega=\omega_1,\tilde{\omega}=\tilde{\omega}_1)=\lambda$ and $\mathbb{P}\times \tilde{\mathbb{P}}(\omega=\omega_2,\tilde{\omega}=\tilde{\omega}_1)=1-\lambda$, where $0<\lambda<1$ is a constant. Then, for any risks $(X,Y)$ with $X(\omega_i)=x_i\ge 0$ and $Y(\omega_i)=y_i$, $i=1,2$, it holds that
$$\mathbb{E}_{\mathbb{P}\times \tilde{\mathbb{P}}}\big[U\big(X, Y,\tilde{\omega}\big)\big]=(x_1^{1/2}+y_1)\lambda+(x_2^{1/2}+y_2)(1-\lambda)$$
and 
$$\mathbb{E}_{\mathbb{P}\times \tilde{\mathbb{P}}}\big[U\big(\mathbb{E}_{\mathbb{P}}[X], Y,\tilde{\omega}\big)\big]=\Big(\big(x_1\lambda+x_2(1-\lambda)\big)^{1/2}+y_1\Big)\lambda+\Big(\big(x_1\lambda+x_2(1-\lambda)\big)^{1/2}+y_2\Big)(1-\lambda),$$
and thus $\mathbb{E}_{\mathbb{P}\times \tilde{\mathbb{P}}}\big[U\big(X, Y,\tilde{\omega}\big)\big]\le\mathbb{E}_{\mathbb{P}\times \tilde{\mathbb{P}}}\big[U\big(\mathbb{E}_{\mathbb{P}}[X], Y,\tilde{\omega}\big)\big]$.  

Summarizing the above results, we provide a counterexample in which the SDBUF $U$ is risk averse to $X$ in the presence of a background risk $Y$, but the SDBUF $U$ is not statewise concave in its first argument.  
\end{example}

\subsection{Uniform bivariate risk aversion for state-dependent utilities $U(x,y,\tilde{\omega})$  }

In this subsection, we examine the economic and mathematical characteristics of statewise conditions under which  the SDBUF $U(x,y,\tilde{\omega})$ exhibits bivariate risk aversion toward $(X,Y)$. To this end, we introduce a new concept termed uniform risk aversion in the presence of a background risk, and provide the equivalent condition for it. 

Consider the product probability space $(\Omega\times \tilde{\Omega}, \mathcal{F}\times \tilde{\mathcal{F}}, \mathbb{P}\times \tilde{\mathbb{P}} )$. Let $\mathcal{Q}$ be the set of all the non-trivial probability measures, that is, all the probability measures $\mathbb{P}\times \tilde{\mathbb{P}}$ such that there exists a set $A\in \mathcal{F}\times \tilde{\mathcal{F}}$ satisfying   $(\mathbb{P}\times \tilde{\mathbb{P}})(A)>0$ and $(\mathbb{P}\times \tilde{\mathbb{P}})(A^c)>0$, where $A^c$ is the complementary set of $A$. 
We begin by formally defining uniform risk aversion in the presence of a background risk for the SDBUF $U(x,y,\tilde{\omega})$. 
This concept draws inspiration from the notation of uniform risk aversion for single-risk introduced by \cite{2021Jarrow}, as well as from the characterization of risk aversion in financial economics \citep{Ingersoll1987}. 

\begin{defn}\label{uniform-risk-aversion-1}A SDBUF   $U(x,y,\tilde{\omega})$ is said to be uniformly risk averse to  $X$ in the presence of a background risk $Y$ if the inequality  $\mathbb{E}_{\mathbb{P}\times \tilde{\mathbb{P}}}\big[U(X, Y,\tilde{\omega})\big]\le \mathbb{E}_{\mathbb{P}\times \tilde{\mathbb{P}}}\big[U(\mathbb{E}_{\mathbb{P}}[X], Y,\tilde{\omega})\big]$ 
	holds for  all $\mathbb{P}\times \tilde{\mathbb{P}}\in \mathcal{Q}$. 
\end{defn}

The uniform risk aversion in the presence of a background risk is a stronger definition than the concept of bivariate risk aversion under state-dependent utilities introduced in Definition \ref{3.1}. Specifically,  for any risks $X$ and $Y$, it requires that the inequality $\mathbb{E}_{\mathbb{P}\times \tilde{\mathbb{P}}}\big[U(X, Y,\tilde{\omega})\big]\le  \mathbb{E}_{\mathbb{P}\times \tilde{\mathbb{P}}}\big[U(\mathbb{E}_{\mathbb{P}}[X], Y,\tilde{\omega})\big]$ holds for all non-trivial probability measures $\mathbb{P}\times \tilde{\mathbb{P}}$. For brevity, when the SDBUF $U(x,y,\tilde{\omega})$ is uniformly risk averse to $X$ in the presence of a background risk $Y$, it is 
also said to be uniformly bivariate risk averse toward  $(X,Y)$.  

Propositions \ref{cor:Uindependent-gamble-sufficient}-\ref{cor:Uindependent-gamble-sufficient++} have established the statewise conditions on the SDBUF  $U(x,y,\tilde{\omega})$ that guarantee risk aversion to  $X$ in the presence of a background risk $Y$  under various correlation structures between the two risks. The following theorem shows that these statewise conditions are equivalent to uniform bivariate risk aversion toward $(X,Y)$.


\begin{thm}\label{thm:increasing-gamble-sufficient}
	Suppose that $U(x,y,\tilde{\omega})$ is a SDBUF with respect to the state space $\tilde{\Omega}$. 
	\begin{itemize}
	\item[($i$)] For any statistically independent risks $(X,Y)$, 
$U(x,y,\tilde{\omega})$ is uniformly risk averse to $X$ in the presence of $Y$ if and only if  $U(x,y,\tilde{\omega})$ is statewise concave in its  first argument $x$.

	\item[($ii$)]	For any risks $(X,Y)\in \mathbb{E}_{\mathbb{P}}D_{N}(X|Y)$, $U(x,y,\tilde{\omega})$ is uniformly risk averse to $X$ in the presence of  $Y$ if and only if   for each state  $\tilde{\omega}\in\tilde{\Omega}$, $U(x,y,\tilde{\omega})$ is concave in $x$  and  
	obeys risk apportionment of order $(1,k)$ for $k\in\{1,\ldots, N\}$.

\item[($iii$)]	For any risks $(X,Y)\in R^{\mathbb{P}}(X|Y)$, $U(x,y,\tilde{\omega})$ is uniformly risk averse to $X$ in the presence of $Y$ if and only if for each $\tilde{\omega}\in\tilde{\Omega}$,  $U(x,y,\tilde{\omega})$ is concave in $x$  and  
obeys risk apportionment of order $(1,1)$.

	\item[($iv$)] For an arbitrary pair of risks $(X,Y)$, $U$ is uniformly risk averse to $X$ in the presence of  $Y$ if and only if for each $\tilde{\omega}\in\tilde{\Omega}$,  $U(x,y,\tilde{\omega})=f(x,\tilde{\omega})+g(y,\tilde{\omega})$, 
where $f(x,\tilde{\omega})$ is statewise concave in $x$.
\end{itemize} 
\end{thm}

For pairs of risks $(X,Y)$ with different forms of correlation,   Theorem  \ref{thm:increasing-gamble-sufficient} establishes the equivalent statewise conditions under which the SDBUF $U(x,y,\tilde{\omega})$ exhibits uniform bivariate risk aversion toward $(X,Y)$. Importantly, 
Theorem  \ref{thm:increasing-gamble-sufficient} combines three  concepts within the theory of risk preferences: ($i$) uniform bivariate risk aversion for state-dependent utilities $U(x,y,\tilde{\omega})$, ($ii$) risk
apportionment as formulated by \cite{2006Putting}, and ($iii$) utility independence \citep{Keeney1976}. Hence, this result is 
not only an extension of existing analyses, but also establishes a unified theoretical framework for these fundamental notations. It thereby offers deeper insight into both the economic interpretation and mathematical characterisation of different statewise conditions. For example, the theorem shows that, for a given SDBUF $U(x,y,\tilde{\omega})$,  the statewise mutual utility independence is equivalent to uniform bivariate risk aversion toward all possible pairs of risks.

Moreover, Theorem  \ref{thm:increasing-gamble-sufficient} highlights a duality relationship between the correlation structure of risks  $(X,Y)$ and the functional form of the SDBUF $U(x,y,\tilde{\omega})$ ensuring uniform bivariate risk aversion. Specifically, for each state $\tilde{\omega}\in\tilde{\Omega}$, the stronger the imposed constraints on the SDBUF $U(x,y,\tilde{\omega})$, the boarder the class 
of admissible correlations between $X$ and $Y$ that allow the preservation of uniform risk aversion to $X$ in the presence of $Y$.

\section{Bivariate risk aversion for dependent lotteries}\label{chaper:uniform}

In the section, we examine Case $II$, where $(\tilde{\Omega}, \tilde{\mathcal{F}}, \tilde{\mathbb{P}} )=(\Omega, \mathcal{F}, \mathbb{P} )$, that is, the state spaces $\Omega$ and $\tilde{\Omega}$ coincide. Under this setting, the risks $X$, $Y$ and the SDBUF $U(x,y,\omega)$ are all defined on the common probability space $(\Omega, \mathcal{F}, \mathbb{P} )$, implying that they share the same source of uncertainty. Within this framework, we identify the conditions under which risk aversion to a primary risk   $X$ is preserved in the presence of a background risk  $Y$ for the SDBUF  $U(x,y,\omega)$.

\subsection{Bivariate risk aversion for state-dependent utilities $U(x,y,\omega)$  }\label{4.1}

We begin by introducing the formal definition of risk aversion in the presence of a background risk for state-dependent utilities $U(x,y,\omega)$. 

\begin{defn}\label{risk-aversion1} A SDBUF  $U(x,y,\omega)$ is said to be risk averse to $X$ in the presence of a background risk $Y$ if the  inequality 
\begin{equation}\label{eq:1}
\mathbb{E}_{\mathbb{P}}\big[U(X, Y,\omega)\big]\le  \mathbb{E}_{\mathbb{P}}\big[U(\mathbb{E}_{\mathbb{P}}[X], Y,\omega)\big]
\end{equation}
holds. 
\end{defn}

The inequality \eqref{eq:1} indicates that when an agent with the SDBUF $U(x, y, \omega)$ is risk averse to $X$ in the presence of a background risk $Y$, he/she prefers  $(\mathbb{E}_{\mathbb{P}}[X],Y)$
to the risks $(X,Y)$. 
For brevity, when the inequality \eqref{eq:1} holds,  the SDBUF $U(x,y,\omega)$ is also said to bivariate risk averse toward  the risks $(X,Y)$.

For the SDBUF $U(x,y,\tilde{\omega})$ being statistically independent of the state space of $(X,Y)$, Propositions \ref{cor:Uindependent-gamble-sufficient}-\ref{cor:Uindependent-gamble-sufficient++} establish the statewise conditions on  $U(x,y,\tilde{\omega})$ that ensure risk aversion to $X$ in the presence of $Y$  under various correlation  structures between the two risks. For example, when $(X,Y)$ exhibit an arbitrary correlation structure,  the statewise mutual utility independence, corresponding to the additive form  $U(x,y,\tilde{\omega})=f(x,\tilde{\omega})+g(y,\tilde{\omega})$, where $f(x,\tilde{\omega})$ is statewise concave in $x$, is sufficient to guarantee bivariate risk aversion toward $(X,Y)$. However, these results do not necessarily hold for the SDBUF $U(x,y,\omega)$. In the following, we clarify this point by taking the additive form of the SDBUF $U(x,y,\omega)$ as an example.


\begin{example}\label{counterpart-add}
 Let $\Omega=\{\omega_1, \omega_2\}$ and $\mathbb{P}(\omega=\omega_1)=\mathbb{P}(\omega= \omega_2)=1/2$. Also, let the SDBUF $U(x,y,\omega)$ be 
 \begin{equation}\label{eq:utilityadd}
 	U(x,y,\omega)=\big(\ln x+y\big) \mathbb{I}_{\{\omega=\omega_1\}}(\omega)+\big(2\ln x+y\big) \mathbb{I}_{\{\omega=\omega_2\}}(\omega).
 \end{equation}

 Note that for each $\omega\in\Omega$, $U(x,y,\omega)$ is concave in its first argument $x$. Take $X(\omega_1)=6$, $X(\omega_2)=10$ and $Y(\omega_i)=y_i$, $i=1,2$. 
 
 	We first calculate $\mathbb{E}_{\mathbb{P}}\big[U(X,Y,\omega) \big]$. From the expression \eqref{eq:utilityadd} of the SDBUF $U$, we have that 
$$\mathbb{E}_{\mathbb{P}}\big[U(X,Y,\omega) \big]= \big(\ln 6+y_1\big)\mathbb{P}(\omega=\omega_1)+ \big(2\ln 10+y_2\big)\mathbb{P}(\omega=\omega_2)=\frac{\ln 6}{2}+\ln 10+\frac{y_1+y_2}{2}.$$
Next, we calculate  $\mathbb{E}_{\mathbb{P}}\big[U(\mathbb{E}_{\mathbb{P}}[X], Y,\omega)\big]$. Noting that   $\mathbb{E}_{\mathbb{P}}\big[X\big]=8$, it yields that 
$$\mathbb{E}_{\mathbb{P}}\big[U(\mathbb{E}_{\mathbb{P}}[X], Y,\omega)\big] = \big(\ln 8+y_1\big)\mathbb{P}(\omega=\omega_1)+ \big(2\ln 8+y_2\big)\mathbb{P}(\omega=\omega_2)
=\frac{3\ln 8}{2}+\frac{y_1+y_2}{2}. $$

 	Then, $\mathbb{E}_{\mathbb{P}}\big[U(X,Y,\omega) \big]>\mathbb{E}_{\mathbb{P}}\big[U(\mathbb{E}_{\mathbb{P}}[X], Y,\omega)\big]$.  Thus, for the SDBUF $U(x,y,\omega)$, the statewise additive form, combined with the statewise concavity in the first argument, is not sufficient to ensure that $U$ exhibits bivariate risk aversion toward $(X,Y)$.
 \end{example}

In the following, we establish a sufficient condition under which, for a general SDBUF $U(x,y,\omega)$ and any pair of risks $(X,Y)$,  the SDBUF $U$ preserves risk aversion toward $X$ in the presence of $Y$. 

\begin{thm}\label{thm:independent-gamble-sufficient}
For any pair of risks $(X,Y)$, if for each state $\omega\in\Omega$,  $U(x,y,\omega)$ is concave in its first argument $x$,  and the covariance $${\rm Cov}_{\mathbb{P}}(X, U^{(1,0)}\big(\mathbb{E}_{\mathbb{P}}[X],Y,\omega)\big)\le 0 $$
	 holds, then $U$ is risk averse to $X$ in the presence of  $Y$. 
\end{thm}

Theorem  \ref{thm:independent-gamble-sufficient} establishes a sufficient condition for the SDBUF $U(x,y,\omega)$ to be bivariate risk aversion toward $(X,Y)$.  Since both the SDBUF $U(x,y,\omega)$ and the risks $(X,Y)$ are defined on the common state space $\Omega$, this sufficient condition  fundamentally depends on the joint functional relationship among $U$, $X$ and $Y$. Even when risks $X$ and $Y$ are statistically independent, or when $U(x,y,\omega)$ satisfies statewise mutual utility independence, the requirement that ${\rm Cov}_{\mathbb{P}}\big(X, U^{(1,0)}(\mathbb{E}_{\mathbb{P}}[X],Y,\omega)\big)\le 0$ generally cannot  
be separated into conditions that depend  solely on the form of   $U(x,y,\omega)$ or solely on the correlation structure of $(X,Y)$. 

\begin{rem} In the single-risk case where only $X$ is considered, \cite{2021Jarrow} derived a similar sufficient condition, namely ${\rm Cov}_{\mathbb{P}}\big(X,  U'(\mathbb{E}_{\mathbb{P}}[X],\omega)\big)\le 0$, for a state-dependent single-attribute utility function $U(x,\omega)$ to exhibit  risk aversion to $X$. Furthermore, they provided an economic interpretation for the  
non-positivity of the covariance ${\rm Cov}_{\mathbb{P}\times \tilde{\mathbb{P}}}(X,$
$ U'(\mathbb{E}_{\mathbb{P}\times \tilde{\mathbb{P}}}[X],\omega))$ within the context of asset pricing. They showed that when the  return on an asset is negatively correlated with the marginal utility of consumption, the asset is perceived as risky, thereby commanding a risk premium. This economic interpretation can be extended naturally to the bivariate setting. Specifically, the  
non-positivity of the covariance ${\rm Cov}_{\mathbb{P}}(X, U^{(1,0)}\big(\mathbb{E}_{\mathbb{P}}[X],Y,\omega)\big)$ can be understood analogously in an asset pricing framework where the agent's utility depends on both wealth $X$ and a background risk $Y$. For a more detail discussion of the economic interpretation for the non-positivity of this covariance,  we refer to \cite{2021Jarrow}.
\end{rem}

\begin{rem}
 The conclusion of Theorem  \ref{thm:independent-gamble-sufficient}  also admits a natural connection to the state-independent utility framework. More specifically, if $U$ is a state-independent, that is, for all $\omega\in\Omega$,  $U(x,y,\omega)=U(x,y)$, then  $U^{(1,0)}(\mathbb{E}_{\mathbb{P}}[X],Y,\omega)=U^{(1,0)}(\mathbb{E}_{\mathbb{P}}[X],Y)$ so that the theorem reduces to the corresponding state-independent case.
	\begin{itemize}
		\item When $X$ and $Y$ are statistically independent, it follows that $X$ and $U^{(1,0)}(\mathbb{E}_{\mathbb{P}}[X],Y)$ are also statistically independent, implying ${\rm Cov}_{\mathbb{P}}\big(X,$
		$U^{(1,0)}(\mathbb{E}_{\mathbb{P}}[X],Y)\big)= 0$. Thus, by  Theorem  \ref{thm:independent-gamble-sufficient}, we know that the concavity of  $U(x,y)$ in $x$  implies  risk aversion to $X$ in the presence of $Y$. 
		
		In contrast, when $X$ and $Y$ are comonotonic risks, the condition that $U^{(1,0)}(x,y)$ is non-increasing in $y$ implies that $X$ and $U^{(1,0)}(\mathbb{E}_{\mathbb{P}}[X],Y)$ are negatively correlated random variables  such that ${\rm Cov}_{\mathbb{P}}(X,$
		$ U^{(1,0)}(\mathbb{E}_{\mathbb{P}}[X], Y))\le 0 $ holds. From the assumption that  $U(x,y)$    
		obeys risk apportionment of order $(1,1)$, we know that  $U^{(1,0)}(x,y)$ is non-increasing in $y$. Thus, it follows from Theorem  \ref{thm:independent-gamble-sufficient}  that  the  concavity of  $U(x,y)$ in $x$, together with the condition  that  $U(x,y)$    
		obeys risk apportionment of order $(1,1)$,  implies risk aversion toward $X$ in the presence of $Y$.
		
		The above results are consistent with Theorems 1 and 2 in \cite{1999On}, which provided the sufficient and necessary conditions for risk aversion to one risk in the presence of a background risk under a state-independent utility framework. 
		
		\item When $(X,Y)\in \mathbb{E}_{\mathbb{P}}D_{N}(X|Y)$, $N\ge1$, following the proof of Proposition \ref{cor:Uindependent-gamble-sufficient+} and using simple manipulation, we know that if $U(x,y)$  is concave in its first argument $x$ and obeys risk apportionment of order $(1,k)$ for $k\in\{1,\ldots, N\}$, then the covariance condition ${\rm Cov}_{\mathbb{P}}(X,$
		$ U^{(1,0)}(\mathbb{E}_{\mathbb{P}}[X], Y))\le 0 $ holds. Consequently,  $U$ exhibits risk aversion to $X$ in the presence of $Y$. This result is consistent with Proposition 3.2 in \cite{2016Risk}, which establishes conditions for risk aversion toward $(X,Y)\in \mathbb{E}_{\mathbb{P}}D_{N}(X|Y)$  under state-independent utilities. 
	\end{itemize}
\end{rem}

In the following example, we show that the converse to Theorem \ref{thm:independent-gamble-sufficient} is not right.

\begin{example}\label{counterpart-2}

Let $\Omega=[0,1]$ and $\mathbb{P}$ be the Lebesgue measure on the unit interval, that is, ${\rm d}\mathbb{P}(\omega)={\rm d}\omega$. 
In this case, $\mathbb{E}_{\mathbb{P}}\big[\omega\big]=1/2$ and $\mathbb{P}(\omega\le \mathbb{E}_{\mathbb{P}}\big[\omega\big])=\mathbb{P}(\omega> \mathbb{E}_{\mathbb{P}}\big[\omega\big])=1/2$. 	 Let the SDBUF  $U(x,y,\omega)$ with respect to $\Omega$ be 
\begin{equation}\label{eq:utility}
	U(x,y,\omega)=2\epsilon x^{\lambda} y^{-1/2} \mathbb{I}_{[0,\mathbb{E}_{\mathbb{P}}[\omega]]}(\omega)+2 x^{\lambda}y^{-1/2}\mathbb{I}_{(\mathbb{E}_{\mathbb{P}}[\omega],1]}(\omega),
\end{equation}
where $0\le \epsilon<1-1/\sqrt{2}$ and $1>\lambda>1/2$. Notice easily that for each $\omega\in\Omega$, $U(x,y,\omega)$ is concave in $x$.

Take  $X(\omega)=Y(\omega)=\omega$. To construct a counterexample, in the following, we first show that $\mathbb{E}_{\mathbb{P}}\big[U(X,Y,\omega) \big]\le  \mathbb{E}_{\mathbb{P}}\big[U(\mathbb{E}_{\mathbb{P}}[X],Y,\omega) \big]$ for risks $X$ and $Y$, that is, $U$ is risk averse to $X$ in the presence of $Y$, and then verify that the covariance ${\rm Cov}_{\mathbb{P}}\big(X, U^{(1,0)}(\mathbb{E}_{\mathbb{P}}[X],Y,\omega)\big)> 0 $ for risks $X$ and $Y$. 

\noindent (i) Show that there exists $\epsilon\in[0,1-1/\sqrt{2})$ such that 
$\mathbb{E}_{\mathbb{P}}\big[U(X,Y,\omega) \big]\le  \mathbb{E}_{\mathbb{P}}\big[U(\mathbb{E}_{\mathbb{P}}[X],Y,\omega) \big]$. 

Substituting the equalities $X(\omega)=Y(\omega)=\omega$ and
$\mathbb{E}_{\mathbb{P}}\big[X\big]=1/2$ into the SDBUF  \eqref{eq:utility}, and then taking the expectations  $\mathbb{E}_{\mathbb{P}}\big[U(\mathbb{E}_{\mathbb{P}}[X],Y,\omega) \big]$ and $\mathbb{E}_{\mathbb{P}}\big[U(X,Y,\omega) \big]$ respectively, it yields that  
\begin{equation*}
\mathbb{E}_{\mathbb{P}}\big[U(\mathbb{E}_{\mathbb{P}}[X],Y,\omega) \big] =2^{1-\lambda}\big(\epsilon  \mathbb{E}_{\mathbb{P}}[\omega^{-1/2} \mathbb{I}_{[0,1/2]}(\omega)]+ \mathbb{E}_{\mathbb{P}}[\omega^{-1/2}\mathbb{I}_{(1/2,1]}(\omega)]\big) =2^{1-\lambda}\big(\epsilon \sqrt{2} + 2-\sqrt{2}\big) \notag
\end{equation*}	
and 
\begin{eqnarray*}
	&\mathbb{E}_{\mathbb{P}}\big[U(X,Y,\omega) \big]&=2\epsilon  \mathbb{E}_{\mathbb{P}}[\omega^{\lambda-1/2} \mathbb{I}_{[0,1/2]}(\omega)]+ 2\mathbb{E}_{\mathbb{P}}[\omega^{\lambda-1/2}\mathbb{I}_{(1/2,1]}(\omega)]\\
	&&=\frac{2\epsilon }{1+\lambda-1/2}2^{-(1+\lambda-1/2)} + \frac{2}{1+\lambda-1/2}(1-2^{-(1+\lambda-1/2)}).
\end{eqnarray*}
Then
$$\mathbb{E}_{\mathbb{P}}\big[U(\mathbb{E}_{\mathbb{P}}[X],Y,\omega) \big]-\mathbb{E}_{\mathbb{P}}\big[U(X,Y,\omega) \big]=\frac{2^{5/2-\lambda}\lambda}{1+2\lambda}\epsilon-\frac{2^{2-\lambda}(2^{\lambda}-1+(\sqrt{2}-2)\lambda)}{1+2\lambda}.$$
Since $\lambda>0$, then $\epsilon>\epsilon_0(\lambda):=\frac{2^{\lambda}-1+(\sqrt{2}-2)\lambda}{\sqrt{2}\lambda}$ implies that  $\mathbb{E}_{\mathbb{P}}\big[U(X,Y,\omega) \big]\le\mathbb{E}_{\mathbb{P}}\big[U(\mathbb{E}_{\mathbb{P}}[X],Y,\omega) \big]$.

Recall that $1/2< \lambda<1$. Hence, it suffices to verify that $\epsilon_0(\lambda)\in(0,1-1/\sqrt{2})$ for all $1/2< \lambda<1$. Note that 
$\epsilon'_0(\lambda)=\frac{1 - 2^{\lambda} + 2^{\lambda} \lambda \ln 2}{\sqrt{2} \lambda^2}$. Denote $f(\lambda):=1 - 2^{\lambda} + 2^{\lambda} \lambda \ln 2$. It is easy to see that  $f'(\lambda)=2^{\lambda}\lambda(\ln 2)^2>0$ for all $1/2< \lambda<1$. Then $f(\lambda)$ is strictly increasing in $(1/2,1)$ such that  $f(\lambda)> f(1/2)=(4-\sqrt{2}+2\sqrt{2}\ln 2)/\sqrt{2}>0$ for all $1/2<\lambda<1$. Recall that $\epsilon'_0(\lambda)=f(\lambda)/\sqrt{2} \lambda^2$. Consequently, $\epsilon'_0(\lambda)>0$ for all $1/2< \lambda<1$. Then $\epsilon_0(\lambda)$ is strictly increasing in $(1/2,1)$, and thus for all $1/2<\lambda<1$, 
$$0<\frac{3\sqrt{2}-4}{\sqrt{2}}=\epsilon_0(\frac{1}{2})<\epsilon_0(\lambda)< \epsilon_0(1)=1-\frac{1}{\sqrt{2}}.$$
	
\noindent (ii) Show that the covariance ${\rm Cov}_{\mathbb{P}}\big(X, U^{(1,0)}(\mathbb{E}_{\mathbb{P}}[X],Y,\omega)\big)> 0 $ for risks $X$ and $Y$ when $\epsilon\in[0,1-1/\sqrt{2})$. 

From the expression \eqref{eq:utility} of the SDBUF, we know that 
\begin{equation*}\label{eq:Dutility}
U^{(1,0)}(x,y,\omega)=2\lambda\epsilon x^{\lambda-1} y^{-1/2} \mathbb{I}_{[0,1/2]}(\omega)+2 \lambda x^{\lambda-1}y^{-1/2}\mathbb{I}_{(1/2,1]}(\omega).
\end{equation*}
Substituting the expressions of risks $X$, $Y$ and $\mathbb{E}_{\mathbb{P}}[X]$ into the above equality and taking expectations on both side of this equality, it yields that 
	\begin{eqnarray*}
	&\mathbb{E}_{\mathbb{P}}\big[U^{(1,0)}(\mathbb{E}_{\mathbb{P}}[X],Y,\omega) \big]& =2^{2-\lambda}\lambda\big(\epsilon  \mathbb{E}_{\mathbb{P}}[\omega^{-1/2} \mathbb{I}_{[0,1/2]}(\omega)]+ \mathbb{E}_{\mathbb{P}}[\omega^{-1/2}\mathbb{I}_{(1/2,1]}(\omega)]\big)\\
	&&=2^{2-\lambda}\lambda\big(\epsilon \int_{0}^{1/2} \omega^{-1/2} {\rm d} \omega + \int_{1/2}^1 \omega^{-1/2} {\rm d} \omega\big) \\ 
	&& =2^{2-\lambda}\lambda\big(\epsilon \sqrt{2} + 2-\sqrt{2}\big).\notag
\end{eqnarray*}	
Also, we have that 
	\begin{eqnarray*}
	&
\mathbb{E}_{\mathbb{P}}\big[XU^{(1,0)}(\mathbb{E}_{\mathbb{P}}[X],Y,\omega) \big]&=2^{2-\lambda}\lambda\big(\epsilon  \mathbb{E}_{\mathbb{P}}[\omega^{1/2} \mathbb{I}_{[0,1/2]}(\omega)]+ \mathbb{E}_{\mathbb{P}}[\omega^{1/2}\mathbb{I}_{(1/2,1]}(\omega)]\big)\\
&&=2^{2-\lambda}\lambda\big(\frac{\epsilon }{3 \sqrt{2}} + \frac{4-\sqrt{2}}{6}\big).
\end{eqnarray*}	
Then
	\begin{eqnarray*}
		&{\rm Cov}_{\mathbb{P}}\big(X, U^{(1,0)}(\mathbb{E}_{\mathbb{P}}[X],Y,\omega)\big)&=\mathbb{E}_{\mathbb{P}}\big[XU^{(1,0)}(\mathbb{E}_{\mathbb{P}}[X],Y,\omega) \big]-\mathbb{E}_{\mathbb{P}}\big[X\big]\mathbb{E}_{\mathbb{P}}\big[U^{(1,0)}(\mathbb{E}_{\mathbb{P}}[X],Y,\omega) \big]\\
		&&=2^{2-\lambda}\lambda\big(-\frac{\epsilon \sqrt{2}}{3 } + \frac{\sqrt{2}-1}{3}\big)> 0
	\end{eqnarray*}
for all $\epsilon\in[0,1-1/\sqrt{2})$.

 Summarizing the above results, we provide a counterexample in which the SDBUF $U(x,y,\omega)$ is risk averse to $X$ in the presence of a background risk $Y$ and for each $\omega\in\Omega$,  $U(x,y,\omega)$ is concave in its first argument $x$, but the covariance ${\rm Cov}_{\mathbb{P}}(X$, $ U^{(1,0)}\big(\mathbb{E}_{\mathbb{P}}[X], 
 Y,\omega)\big)> 0 $.

\end{example}

\subsection{Uniform bivariate risk aversion for state-dependent utilities $U(x,y,\omega)$   }

Given the probability space $(\Omega, \mathcal{F}, \mathbb{P})$, let $\mathcal{P}$ be the set of all non-trivial probability measures, that is, all the probability measures $\mathbb{P}$ satisfying that there exists a set $A\in \mathcal{F}$ such that  $\mathbb{P}(A)>0$ and $\mathbb{P}(A^c)>0$. In the following, we introduce the definition of uniform risk aversion in the presence of a background risk for state-dependent utilities $U(x,y,\omega)$.

\begin{defn}\label{uniform-risk-aversion} A SDBUF  $U(x,y,\omega)$ is said to be uniformly risk averse to one risk $X$ in the presence of a background risk $Y$ if the inequality $\mathbb{E}_{\mathbb{P}}\big[U(X, Y,\omega)\big]\le  \mathbb{E}_{\mathbb{P}}\big[U(\mathbb{E}_{\mathbb{P}}[X], Y,\omega)\big]$ 
	holds for  all $\mathbb{P}\in \mathcal{P}$. 
\end{defn}

 For simplicity of terminology, when the SDBUF $U$ is uniformly risk averse to $X$ in the presence of a background risk $Y$, it
is also said to be uniformly bivariate risk averse to risks $(X,Y)$. The uniform bivariate risk aversion for $U(x,y,\omega)$ is a stronger definition than bivariate risk aversion introduced in Definition  \ref{risk-aversion1}. Specifically, it requires that for any pair of risks $(X,Y)$,  the inequality $\mathbb{E}_{\mathbb{P}}\big[U(X, Y,\omega)\big]\le \mathbb{E}_{\mathbb{P}}\big[U(\mathbb{E}_{\mathbb{P}}[X], Y,\omega)\big]$ holds for all non-trivial probability measures $\mathbb{P}\in \mathcal{P}$.  

Theorem  \ref{thm:increasing-gamble-sufficient} focuses on the case in which the SDBUF $U(x,y,\tilde{\omega})$ is statistically independent of risks $X$ and $Y$, and provide equivalent conditions under which   $U(x,y,\tilde{\omega})$ is uniformly risk averse to $X$ in the presence of $Y$ for various forms of correlation  between risks $X$ and $Y$. In the following, we extend this analysis to  the SDBUF $U(x,y,\omega)$, and present the equivalent conditions  for $U(x,y,\omega)$ to be uniformly bivariate risk averse.

First, for any pair of risks $(X,Y)\in \mathbb{E}_{\mathbb{P}}D_{N}(X|Y)$, we establish the equivalent condition for  $U(x,y,\omega)$ to be  uniformly risk averse to $X$ in the presence of $Y$ by using the notation of utility dominant attribute.

\begin{thm}\label{thm:expectation-dependent-gamble-sufficient-necessary}
Let  $U(x,y,\omega)$ be a SDBUF, and for each  $\omega\in\Omega$, $x$ be a utility dominant attribute. Suppose that for each $\omega\in\Omega$,  $U(x,y,\omega)$ is concave in its first argument  $x$ and 
obeys risk apportionment of order $(1,k)$ for $k\in\{1,\ldots, N\}$. Then, for any pair of risks $(X,Y)\in \mathbb{E}_{\mathbb{P}}D_{N}(X|Y)$, the following statements are equivalent:
	\begin{itemize}
		\item[(E1)]  $U(x,y,\omega)$ is uniformly risk averse to $X$ in the presence of  $Y$.  
		
		\item[(E2)]  The covariance condition that ${\rm Cov}_{\mathbb{P}}\big(X,U^{(1,0)}(\mathbb{E}_{\mathbb{P}}[X],Y,\omega)\big)\le 0$ holds for all $\mathbb{P}\in \mathcal{P}$. 
		
		\item[(E3)] The bivariate utility function $U(x,y,\omega)$ is state-independent.
	\end{itemize}	
	
\end{thm}

For any pair of risks $(X,Y)\in \mathbb{E}_{\mathbb{P}}D_{N}(X|Y)$, Theorem  \ref{thm:expectation-dependent-gamble-sufficient-necessary} not only establishes a converse to Theorem  \ref{thm:independent-gamble-sufficient}, but also presents an  equivalence between bivariate uniform risk aversion and the state-independent  preference toward the risks $(X,Y)$, thereby offering a novel characterization of   state-independent preferences toward two risks with the  $N$th-degree PED.

 Similar to Theorem \ref{thm:expectation-dependent-gamble-sufficient-necessary}, the following theorem presents the equivalent conditions under which $U(x,y,\omega)$ exhibits uniform risk aversion to $X$ in the presence of $Y$ for different correlation structures: (i) a pair of risks $(X,Y)$ with arbitrary correlation structures, (ii) statistically independent risks, and (iii) pairs satisfying $(X,Y)\in R^{\mathbb{P}}(X|Y)$. Since the proof is similar to that of Theorem \ref{thm:expectation-dependent-gamble-sufficient-necessary}, we omit it.  

\begin{thm}\label{thm:independent-gamble-sufficient-necessary}
Let  $U(x,y,\omega)$ be a SDBUF and for each  $\omega\in\Omega$, $x$ be a utility dominant attribute.  
	\begin{itemize}
			\item[($i$)] For an arbitrary  pair of risks  $(X,Y)$, the condition that for each $\omega\in \Omega$,  $U(x,y,\omega)=f(x,\omega)+g( y,\omega)$, 
	where $f(x,\omega)$ is statewise concave in $x$,  guarantees the equivalence among  the statements (E1),  (E2) and (E3).  
	
		\item[($ii$)] For all statistically independent risks  $X$ and $Y$, the statewise concavity of $U(x,y,\omega)$ in $x$ guarantees the equivalence among  the statements (E1),  (E2) and (E3).
		
	    \item[($iii$)] For all pairs of risks  $(X,Y)\in R^{\mathbb{P}}(X|Y)$, the conditions that for each $\omega\in \Omega$, $U(x,y,\omega)$ is concave in its first argument  $x$  and  
	    obeys risk apportionment of order $(1,1)$, guarantee the equivalence among  the statements (E1),  (E2) and (E3).
	\end{itemize}	

\end{thm}

Theorem  \ref{thm:independent-gamble-sufficient-necessary} establishes an equivalence between the uniform bivariate risk aversion to statistically independent risks $(X,Y)$ and the state-independent preference toward the pair of risks $(X,Y)$. Moreover, this theorem extends this equivalence to the case of arbitrary pairs of risks $(X,Y)$, thereby linking uniform bivariate risk aversion to state-independent preferences toward risks with arbitrary correlation structures. Consequently, this theorem provides new insights into the characterization of state-independent  preferences involving two risks under different correlation structures.

\section{Applications}\label{applications}

\subsection{On the demand for fair insurance contracts under state-dependent setting}

We consider a DM with a state-dependent single-attribute utility  function $V(.,\tilde{\omega})$ defined with respect to the state space of nature $\tilde{\Omega}$, who is exposed to two potential sources of monetary losses: $L$, representing losses to buildings, and $H$, representing losses to building  profits. It is  assumed that only  the loss $L$ is insured. Let $\alpha\in[0, 1]$ denote the proportion of the building loss $L$ that the DM chooses to   insure.  A fair insurance contract against the loss $L$ is offered to the DM, with the premium determined as $\alpha \mathbb{E}_{\mathbb{P}}[L]$.  The  payoff of the DM  is
\begin{equation*}
	P(\alpha)=w-(1-\alpha)L-\alpha \mathbb{E}_{\mathbb{P}}[L]-H,
\end{equation*}
where $w$ is DM's initial wealth. If the DM chooses to purchase full insurance, that is, $\alpha= 1$, his/her resulting payoff is given by  $P(1)=w- \mathbb{E}_{\mathbb{P}}[L]-H$. Notice also that
$\mathbb{E}_{\mathbb{P}}[L]= \mathbb{E}_{\mathbb{P}}\big[(1-\alpha)L+\alpha \mathbb{E}_{\mathbb{P}}[L]\big]$. Hence, $\alpha= 1$ is the optimal  choice 
if and only if for all $\alpha\in[0, 1]$,
	\begin{eqnarray}
	\mathbb{E}_{\mathbb{P}\times \tilde{\mathbb{P}}}\big[V\big(w-(1-\alpha)L-\alpha \mathbb{E}_{\mathbb{P}}[L]-H,\tilde{\omega}\big)\big]&\le & \mathbb{E}_{\mathbb{P}\times \tilde{\mathbb{P}}}\big[V\big(	P(1),\tilde{\omega}\big)\big]=\mathbb{E}_{\mathbb{P}\times \tilde{\mathbb{P}}}\big[V\big(	w- \mathbb{E}_{\mathbb{P}}[L]-H,\tilde{\omega}\big)\big]\nonumber\nonumber\\
	&=&\mathbb{E}_{\mathbb{P}\times \tilde{\mathbb{P}}}\big[V\big(	w-\mathbb{E}_{\mathbb{P}}[(1-\alpha)L-\alpha \mathbb{E}_{\mathbb{P}}[L]]-H,\tilde{\omega}\big)\big].\nonumber
\end{eqnarray}

The empirical findings reported in \cite{2016Risk} suggest that losses to buildings exhibit 1st-degree PED on the losses to profits generated by these buildings. From Proposition \ref{cor:Uindependent-gamble-sufficient+}, 
 we obtain the following result for 1st-degree PED losses $L$ and $H$ by letting 
 $U(x, y, \tilde{\omega}) = V(x+y,\tilde{\omega})$. It is worth noting that similar results can be obtained for losses  $L$ and $H$ exhibiting other types of correlation structures.
 
 \begin{prop}\label{insurance} For all losses $(-L,-H)\in\mathbb{E}_{\mathbb{P}}D_1(-L|-H) $, if a DM is risk averse under state-dependent utilities, that is, his/her utility function  $V(.,\tilde{\omega})$ is statewise concave, then full insurance  ($\alpha=1$) is the optimal choice. 
 \end{prop}
 	
According to Proposition \ref{insurance},  when the DMs' utilities are state-dependent and the loss $L$ exhibits 1st-degree PED on $H$, the statewise risk aversion alone suffices to induce the DM to purchase full insurance. 
 	
 \subsection{Optimal prevention	within a two-period model with state-dependent utilities}
 
 In this subsection, we extend the state-dependent analysis to the  optimal prevention within a two-period framework. Consider a DM who lives for two periods, denoted as period 0 and period 1. The
 DM has a two-attribute utility function $U_0(w,h)$ in period 0 and a SDBUF, $U_1(w,h,\tilde{\omega})$ in period 1, defined
 over health status $h$ and wealth level $w$. It is assumed that 
 that $U_0(w,h)$ is non-decreasing in its first argument $w$, and $U_1(w,h,\tilde{\omega})$ is statewise non-decreasing in the wealth  argument $w$. At the beginning of period 0,  the DM knows his/her initial wealth 
 and health status, denoted by $w_0$ and $h_0$, respectively. However, he/she faces uncertainty about utility, wealth, and health status in period 1, represented by a SDBUF $U_1(w,h,\tilde{\omega})$, an  uncertain wealth
 level $W$, and a random health variable $H$.  Specifically, in period 1, the DM's utility $U_1(w,h,\tilde{\omega})$ is state-dependent with respect to the state space of nature $\tilde{\Omega}$, and the DM's wealth level is either $w_1$ in the no-loss case or  $w_1 + \alpha W + (1-\alpha)\mathbb{E}_{\mathbb{P}}[W]$ in the loss case. Here, $\alpha\in(0, 1]$
 measures the degree of exposure to the uncertain loss, which is represented 
 by the negative random variable $W$. The joint distribution of the  random variables $W$ and $H$ is specified by a known CDF,  $F^{\mathbb{P}}(w,h)$, defined over the support 
 $[\underline{w},\overline{w}]\times[\underline{h},\overline{h}]$, where $-w_1 < \underline{w}<\overline{w} < 0$ and $0 < \underline{h}<\overline{h}$.
 
The DM can allocate an amount  $e$ in period 0 toward prevention activities so as to influence the probability of experiencing the loss 
 in period 1. For all $e\in [0,w_0]$, this probability is denoted by $p(e)\in(0, 1)$. 
 Prevention is assumed to be effective such that $p'(e)\le 0$ for all $e\in [0,w_0]$. For brevity, the DM's subjective discount factor is normalized to unity. Consequently, the DM's two-period expected utility function can be expressed by
 \begin{equation}\label{P1+}V(e)=U_0(w_0-e,h_0)+p(e)\mathbb{E}_{\mathbb{P}\times \tilde{\mathbb{P}}}\big[U_1\big(	w_1+\alpha W+(1-\alpha)\mathbb{E}_{\mathbb{P}}[W],H,\tilde{\omega}\big)\big]+\big(1-p(e)\big)\mathbb{E}_{\mathbb{P}\times \tilde{\mathbb{P}}}\big[U_1(w_1,H,\tilde{\omega})\big]. \end{equation}
 	
Assume that there exists an interior optimal solution. The DM's optimal
prevention level, denoted by $e_V$, satisfies the first-order
condition $V'(e_V) = 0$, which can be expressed as follows: 	
 \begin{equation}\label{P1}
 -p'(e_V)\Big(\mathbb{E}_{\mathbb{P}\times \tilde{\mathbb{P}}}\big[U_1(w_1,H,\tilde{\omega})\big]-
 \mathbb{E}_{\mathbb{P}\times \tilde{\mathbb{P}}}\big[U_1\big(	w_1+\alpha W+(1-\alpha)\mathbb{E}_{\mathbb{P}}[W],H,\tilde{\omega}\big)\big]\Big)=U_0^{(1,0)}(w_0-e_V,h_0). 
 \end{equation}
 
 The right-hand
 side of Equation \eqref{P1} represents the marginal cost of prevention, corresponding to 
 the marginal utility in period 0 resulting from the expenditure $e_V$ on
 prevention activities. The left-hand side is the marginal benefit of prevention, measured by the gain in the expected utility resulting from a reduction in
 the probability of experiencing the loss case in period 1.

 At the
 optimal prevention level $e_V$, the DM equilibrates these two effects, ensuring that the marginal cost of
 prevention equals to its marginal benefit.
 
 As a benchmark, let $\tilde{e}_V$ denote the optimal level of prevention in the absence of risk exposure, that is, in the case  $\alpha=0$. In this case, the loss is fixed in the loss case  and equal
 to $\mathbb{E}_{\mathbb{P}}[W]<0$. According, $\tilde{e}_V$ is determined as the unique solution to the following 
equation:
  \begin{equation}\label{P2}
 	-p'(\tilde{e}_V)\Big(\mathbb{E}_{\mathbb{P}\times \tilde{\mathbb{P}}}\big[U_1(w_1,H,\tilde{\omega})\big]-
 	\mathbb{E}_{\mathbb{P}\times \tilde{\mathbb{P}}}\big[U_1\big(	w_1+\mathbb{E}_{\mathbb{P}}[W],H,\tilde{\omega}\big)\big]\Big)=U_0^{(1,0)}(w_0-\tilde{e}_V,h_0). 
 \end{equation}

Taking the first derivation of $V(e)$ in \eqref{P1+} with respect to the prevention level $e$ and subsequently evaluating it at $e = \tilde{e}_V$ gives 
  \begin{equation*}
 V'(\tilde{e}_V)	=p'(\tilde{e}_V)\Big(\mathbb{E}_{\mathbb{P}\times \tilde{\mathbb{P}}}\big[U_1\big(	w_1+\alpha W+(1-\alpha)\mathbb{E}_{\mathbb{P}}[W],H,\tilde{\omega}\big)\big]-
 \mathbb{E}_{\mathbb{P}\times \tilde{\mathbb{P}}}\big[U_1(w_1,H,\tilde{\omega})\big]\Big)-U_0^{(1,0)}(w_0-\tilde{e}_V,h_0) 
 \end{equation*}
 	\begin{equation}\label{P3}
=p'(\tilde{e}_V)\Big( \mathbb{E}_{\mathbb{P}\times \tilde{\mathbb{P}}}\big[U_1\big(	w_1+\alpha W+(1-\alpha)\mathbb{E}_{\mathbb{P}}[W],H,\tilde{\omega}\big)\big]-\mathbb{E}_{\mathbb{P}\times \tilde{\mathbb{P}}}\big[U_1(w_1+\mathbb{E}_{\mathbb{P}}[W],H,\tilde{\omega})\big]\Big),\ \ \ \ \ \ \ \  
 \end{equation} 
where the second equality follows from Eq. \eqref{P2}. Since $p'(e)\le 0$ 
for all  $e\in [0,w_0]$, it follows from Eqs. \eqref{P1} and \eqref{P3} that
$e_V\ge\tilde{e}_V$ for all $\alpha\in(0, 1]$ if and only if the inequality 
$$\mathbb{E}_{\mathbb{P}\times \tilde{\mathbb{P}}}\big[U_1\big(	w_1+\alpha W+(1-\alpha)\mathbb{E}_{\mathbb{P}}[W],H,\tilde{\omega}\big)\big]\le\mathbb{E}_{\mathbb{P}\times \tilde{\mathbb{P}}}\big[U_1(w_1+\mathbb{E}_{\mathbb{P}}[W],H,\tilde{\omega})\big] $$
holds for all $\alpha\in(0,1]$. Noting that $\mathbb{E}_{\mathbb{P}}\big[\alpha W+(1-\alpha)\mathbb{E}_{\mathbb{P}}[W]\big]=\mathbb{E}_{\mathbb{P}}[W]$ and using Proposition \ref{cor:Uindependent-gamble-sufficient+},  
we obtain the following
proposition.

 \begin{prop}\label{prevention} When facing any pair of uncertain outcomes $(W, H)\in \mathbb{E}_{\mathbb{P}}D_{N}(W|H)$, if for each state  $\tilde{\omega}\in\tilde{\Omega}$, a DM's utility  $U_1(w,h,\tilde{\omega})$ is concave in $w$  and  
 	obeys risk apportionment of order $(1,k)$ for $k\in\{1,\ldots, N\}$, then the DM will optimally spend more on
 		prevention activities, that is, $e_V\ge\tilde{e}_V$ for all $\alpha\in(0, 1]$.  
\end{prop}

It is worth noting that similar results can be obtained for  $W$ and $H$ exhibiting other types of correlation structures.
 	
\section{Conclusions}\label{sec:Conclusion}

Risk aversion under state-dependent utilities plays an important role in a range of economic areas, including insurance theory, financial economics and behavioral economics. So far, the majority of studies in these areas have been conducted within the framework of
a single-attribute utility function, implicitly assuming that  the DM has a univariate objective. This assumption constrains the analysis to the single risk or multiple risks expressed in the same dimension.

The purpose of this paper was to extend the discussion to the preservation of risk aversion to a primarily risk in the
presence of a background risk  under state-dependent utilities, specifically in a multidimensional setting. Two cases were considered: ($I$) the two  risks are independent of the state space of  SDBUF $U(x,y,\tilde{\omega})$, and ($II$) the state space of SDBUF $U(x,y, \omega)$ coincides with that of risks under consideration. We  investigated the relationships among risk aversion with two risks, the functional form of the SDBUF, and the degree of correlation between these risks.

In Case $I$, we showed that the conditions under which  $U(x,y,\tilde{\omega})$ exhibits bivariate risk aversion vary substantially depending on the correlation between risks $X$ and $Y$. More specifically, ($i$) for an arbitrary pair of risks $(X,Y)$,  statewise mutual utility independence and concavity  of $U(x,y,\tilde{\omega})$ in the first argument $x$ are sufficient for bivariate risk aversion,  ($ii$) for PRD risks $(X,Y)$, including bivariate Bernoulli and bivariate normal distributions, statewise concavity in the first argument combined with risk apportionment of order (1,1) is required,  ($iii$) for $N$th-degree PED risks $(X,Y)$,   statewise concavity in the first argument together with  risk apportionment of order $(1,k)$ for $k\in\{1,\ldots,N \}$ is required, and ($iv$) for statistically  independent risks, statewise concavity in the first argument of $U(x,y,\tilde{\omega})$ alone is sufficient. These findings reveal 
 a duality between the correlation of random vector $(X,Y)$
and the functional form of the SDBUF $U(x,y,\tilde{\omega})$. 

Based on this insight, we introduced a new definition of bivariate risk aversion, termed uniform risk aversion in the presence of a background risk (or uniform bivariate risk aversion). We further verified that the above established conditions ($i$)-($iv$)  on the SDBUF $U(x,y,\tilde{\omega})$ are equivalent to uniform bivariate risk aversion  under these corresponding correlation structures between  risks $X$ and $Y$. The obtained results  combined these concepts: ($a$) uniform bivariate risk aversion for state-dependent utilities $U(x,y,\tilde{\omega})$, ($b$) risk
apportionment, as formulated by \cite{2006Putting}, and ($c$) utility independence \citep{Keeney1976}. Thus, the economic interpretations of the established conditions ($i$)-($iv$) on the SDBUF $U(x,y,\tilde{\omega})$ were clarified, respectively. 

In Case $II$, where the risks $(X,Y)$ share the same state space as the utility function $U(x,y, \omega)$, the conditions ($i$)-($iv$) established in Case $I$ were shown to be no longer applicable. For this scenario, we propose a sufficient condition guaranteeing the preservation of  risk aversion  toward $X$  in the presence of $Y$. Furthermore, we showed that when the SDBUF $U(x,y, \omega)$ has a utility dominant attribute $x$, the aforementioned conditions ($i$)-($iv$) on  $U(x,y, \omega)$  plus a covariance condition ${\rm Cov}_{\mathbb{P}}(X,U^{(1,0)}\big(\mathbb{E}_{\mathbb{P}}[X],Y,\omega)\big)\le 0$ are equivalent to  the uniform bivariate risk aversion toward $(X,Y)$ under the corresponding correlation between risks $X$ and $Y$.  These findings provide some new theoretical insights into the state-dependent preferences with two correlated risks, deepening our understanding of how correlation influences the preservation of risk aversion toward one risk in the presence of another risk under state-dependent utilities.

Finally, we
offered two applications in the contexts of  fair insurance contracts and optimal prevention within a two-period model to illustrate the usefulness of the proposed concept of bivariate risk aversion under state-dependent utilities. In the demand for fair insurance contract, it was shown that when a DM faces both an insured loss and an uninsured loss, and these losses are 1st-degree PED, the statewise risk aversion alone is sufficient to ensure the optimality of purchasing full insurance. In a two-period optimal prevention model, it was shown that if a DM is bivariate risk averse toward a pair of uncertain variables (an uncertain loss and a health risk), then he/she optimally allocates more on prevention activities.

\section*{Appendix}
\appendix
\section{Some proofs}\label{Appendix:A}
\subsection{Proof of Proposition \ref{cor:Uindependent-gamble-sufficient}}

	If $U(x,y,\tilde{\omega})=f(x,\tilde{\omega})+g(y,\tilde{\omega})$, then 
	\begin{eqnarray}
		\mathbb{E}_{\mathbb{P}\times \tilde{\mathbb{P}}}\big[U\big(X, Y,\tilde{\omega}\big)\big]&=&\mathbb{E}_{\mathbb{P}\times \tilde{\mathbb{P}}}\big[f\big(X,\tilde{\omega}\big)\big]+\mathbb{E}_{\mathbb{P}\times \tilde{\mathbb{P}}}\big[g\big(Y,\tilde{\omega}\big)\big]\nonumber\nonumber\\
		&=&\mathbb{E}_{\mathbb{P}\times \tilde{\mathbb{P}}}\big[\mathbb{E}_{\mathbb{P}\times \tilde{\mathbb{P}}}[f\big(X,\tilde{\omega}\big)|\tilde{\mathcal{F}}]\big]+\mathbb{E}_{\mathbb{P}\times \tilde{\mathbb{P}}}\big[g\big(Y,\tilde{\omega}\big)\big]\nonumber\\
		&=&\mathbb{E}_{\mathbb{P}\times \tilde{\mathbb{P}}}\big[\mathbb{E}_{\mathbb{P}}[f\big(X,\tilde{\omega}\big)]\big]+\mathbb{E}_{\mathbb{P}\times \tilde{\mathbb{P}}}\big[g\big(Y,\tilde{\omega}\big)\big]\nonumber\\
		&\le&\mathbb{E}_{\mathbb{P}\times \tilde{\mathbb{P}}}\big[f\big(\mathbb{E}_{\mathbb{P}}[X],\tilde{\omega}\big)\big]+\mathbb{E}_{\mathbb{P}\times \tilde{\mathbb{P}}}\big[g\big(Y,\tilde{\omega}\big)\big]\nonumber\\
		&=&\mathbb{E}_{\mathbb{P}\times \tilde{\mathbb{P}}}\big[U\big(\mathbb{E}_{\mathbb{P}}[X], Y,\tilde{\omega}\big)\big],
	\end{eqnarray}
	where the second equality holds due to the tower rule for conditional expectations, the third equality follows from the statistical independence of $X$ from $\tilde{\mathcal{F}}$, and the inequality follows from the statewise concavity of $f(x,\omega)$ and Jensen's inequality. Thus, for an arbitrary pair of risks $(X,Y)$, $U$ is risk averse to  $X$ in the presence of $Y$.  

\subsection{Proof of Proposition \ref{cor:Uindependent-gamble-sufficient+}}

	We only verify the statement ($ii$), since the statement ($i$)  can be proved by similar arguments. 
	
	Since $U(x,y,\tilde{\omega})$ is statewise concave in its first argument $x$, then for each state $\tilde{\omega}\in\tilde{\Omega}$, it holds that 
	$$U(X,Y,\tilde{\omega})-U(\mathbb{E}_{\mathbb{P}}[X], Y,\tilde{\omega})\le U^{(1,0)}(\mathbb{E}_{\mathbb{P}}[X], Y,\tilde{\omega}) (X-\mathbb{E}_{ \mathbb{P}}[X]). 
	$$
	Taking expectations on both side of the above inequality, it yields that
	\begin{equation}\label{eq:ex+}
		\mathbb{E}_{\mathbb{P}\times\tilde{\mathbb{P}}}\big[U(X,Y,\tilde{\omega}) \big]-\mathbb{E}_{\mathbb{P}\times\tilde{\mathbb{P}}}\big[U(\mathbb{E}_{\mathbb{P}}[X],Y,\tilde{\omega}) \big]\le \mathbb{E}_{\mathbb{P}\times\tilde{\mathbb{P}}}\big[U^{(1,0)}(\mathbb{E}_{\mathbb{P}}[X], Y,\tilde{\omega}) (X-\mathbb{E}_{\mathbb{P}}[X]) \big].
	\end{equation}
	Simple manipulation yields 
	\begin{eqnarray}
		&&\ \ \ \ \mathbb{E}_{\mathbb{P}\times\tilde{\mathbb{P}}}\big[U^{(1,0)}(\mathbb{E}_{\mathbb{P}}[X], Y,\tilde{\omega}) (X-\mathbb{E}_{\mathbb{P}}[X]) \big]\nonumber\\
		&&=\mathbb{E}_{\mathbb{P}\times \tilde{\mathbb{P}}}\big[XU^{(1,0)}(\mathbb{E}_{\mathbb{P}}[X],Y,\tilde{\omega}) \big]-\mathbb{E}_{\mathbb{P}}\big[X\big]\mathbb{E}_{\tilde{\mathbb{P}}\times\mathbb{P}}\big[U^{(1,0)}(\mathbb{E}_{\mathbb{P}}[X],Y,\tilde{\omega}) \big]\nonumber\\
		&&=\mathbb{E}_{\mathbb{P}\times\tilde{\mathbb{P}}}\Big[\mathbb{E}_{\mathbb{P}\times\tilde{\mathbb{P}}}\big[XU^{(1,0)}(\mathbb{E}_{\mathbb{P}}[X],Y,\tilde{\omega}) |\tilde{\mathcal{F}}\big]\Big]-\mathbb{E}_{\mathbb{P}}\big[X\big]\mathbb{E}_{\tilde{\mathbb{P}}\times\mathbb{P}}\Big[\mathbb{E}_{\tilde{\mathbb{P}}\times\mathbb{P}}\big[U^{(1,0)}(\mathbb{E}_{\mathbb{P}}[X],Y,\tilde{\omega}) |\tilde{\mathcal{F}}\big]\Big]\nonumber\\
		&&=\mathbb{E}_{\mathbb{P}\times\tilde{\mathbb{P}}}\Big[\mathbb{E}_{\mathbb{P}}\big[XU^{(1,0)}(\mathbb{E}_{\mathbb{P}}[X],Y,\tilde{\omega}) \big]\Big]-\mathbb{E}_{\mathbb{P}}\big[X\big]\mathbb{E}_{\tilde{\mathbb{P}}\times\mathbb{P}}\Big[\mathbb{E}_{\mathbb{P}}\big[U^{(1,0)}(\mathbb{E}_{\mathbb{P}}[X],Y,\tilde{\omega}) \big]\Big]\ \ \ \ \ \ \ \ \ \ \ \ \ \ \ \ \ \ \nonumber\\
		&&=\mathbb{E}_{\mathbb{P}\times\tilde{\mathbb{P}}}\Big[\mathbb{E}_{\mathbb{P}}\big[XU^{(1,0)}(\mathbb{E}_{\mathbb{P}}[X],Y,\tilde{\omega}) \big]-\mathbb{E}_{\mathbb{P}}\big[X\big]\mathbb{E}_{\mathbb{P}}\big[U^{(1,0)}(\mathbb{E}_{\mathbb{P}}[X],Y,\tilde{\omega}) \big]\Big]\nonumber\\
		&&=\mathbb{E}_{\mathbb{P}\times\tilde{\mathbb{P}}}\big[{\rm Cov}_{\mathbb{P}}\big(X, U^{(1,0)}(\mathbb{E}_{\mathbb{P}}[X],Y,\tilde{\omega})\big)\big]\nonumber\\
		&&=\mathbb{E}_{\tilde{\mathbb{P}}}\big[{\rm Cov}_{\mathbb{P}}\big(X, U^{(1,0)}(\mathbb{E}_{\mathbb{P}}[X],Y,\tilde{\omega})\big)\big],\label{eq:sufficient-1}
	\end{eqnarray}
	where the second equality holds due to the tower rule for conditional expectations, and the third equality as well as the last equality follow from the statistical independence of $X,Y$ from $\tilde{\mathcal{F}}$.

	From Theorem 1 in \cite{2002On}, we know that for each state $\tilde{\omega}\in\tilde{\Omega}$, 
	\begin{eqnarray}
		&
		{\rm Cov}_{ \mathbb{P}}\big(X,U^{(1,0)}(\mathbb{E}_{\mathbb{P}}[X],Y,\tilde{\omega})\big)&=\int_{\underline{y}}^{\overline{y}}\int_{\underline{x}}^{\overline{x}}U^{(1,1)}(\mathbb{E}_{\mathbb{P}}[X],y,\tilde{\omega})\big(F^{\mathbb{P}}(x,y)-F^{\mathbb{P}}_X(x)F^{\mathbb{P}}_Y(y)\big){\rm d} x {\rm d} y\nonumber\\
		&&=\int_{\underline{y}}^{\overline{y}}U^{(1,1)}(\mathbb{E}_{\mathbb{P}}[X],y,\tilde{\omega})\mathbb{E}_{ \mathbb{P}}D_{1}(X|y)F^{\mathbb{P}}_Y(y) {\rm d} y. \label{eq:sufficient-first1}
	\end{eqnarray}
	By integrating \eqref{eq:sufficient-first1} by parts repeatedly, it yields that for each $\tilde{\omega}\in\tilde{\Omega}$, 
	\begin{eqnarray}
		&
		{\rm Cov}_{ \mathbb{P}}\big(X,U^{(1,0)}(\mathbb{E}_{\mathbb{P}}[X],Y,\tilde{\omega})\big)&=\sum_{k=2}^N(-1)^kU^{(1,k-1)}(\mathbb{E}_{\mathbb{P}}[X],\overline{y},\tilde{\omega})\mathbb{E}_{ \mathbb{P}}D_{k}(X|\overline{y})\nonumber\\
		&&\ \ +\int_{\underline{y}}^{\overline{y}}(-1)^{1+N} U^{(1,N)}(\mathbb{E}_{\mathbb{P}}[X],y,\tilde{\omega})\mathbb{E}_{\mathbb{P}}D_{N}(X|y) {\rm d} y,\ {\rm for}\ N\ge 1, \label{eq:sufficient-first2}
	\end{eqnarray}	
	where $\sum\limits_{k=2}^1(.)=0$. For $(X,Y)\in \mathbb{E}_{ \mathbb{P}}D_{N}(X|Y)$, we know that for each $k\in\{2,\ldots,N\}$,  $\mathbb{E}_{ \mathbb{P}}D_{k}(X|\overline{y})\ge 0$ and 
	for any $y\in[\underline{y},\overline{y}]$, $\mathbb{E}_{ \mathbb{P}}D_{N}(X|y)\ge 0$. Also, for each $\tilde{\omega}\in\tilde{\Omega}$, since  $U(x,y,\tilde{\omega})$ 
	obeys risk apportionment of order $(1,k)$ for $k\in\{1,\ldots, N\}$, then
	$(-1)^{1+k}U^{(1,k)}(x,y,\tilde{\omega})\le 0\ \ {\rm for\ any}\ k=1,\ldots, N$, and thus by \eqref{eq:sufficient-first2} we have that for each $\tilde{\omega}\in\tilde{\Omega}$, ${\rm Cov}_{ \mathbb{P}}\big(X,U^{(1,0)}(\mathbb{E}_{\mathbb{P}}[X],$
	$Y,\tilde{\omega})\big)\le 0$. Combining this result with \eqref{eq:ex+} and \eqref{eq:sufficient-1}, it yields that  
	$\mathbb{E}_{\mathbb{P}\times\tilde{\mathbb{P}}}\big[U(X,Y,\tilde{\omega}) \big]-\mathbb{E}_{\mathbb{P}\times\tilde{\mathbb{P}}}\big[U(\mathbb{E}_{\mathbb{P}}[X],Y,\tilde{\omega}) \big]\le 0$. Thus, for any risks $(X,Y)\in \mathbb{E}_{\mathbb{P}}D_{N}(X|Y)$, $U$ is risk averse to $X$ in the presence of  $Y$.    	

\subsection{Proof of Proposition \ref{cor:Uindependent-gamble-sufficient++}}
	Since $U(x,y,\tilde{\omega})$ is statewise concave in its first argument $x$, then for statistically independent risks $X$ and $Y$, it holds that
	\begin{eqnarray}
		\mathbb{E}_{\mathbb{P}\times \tilde{\mathbb{P}}}\big[U\big(X, Y,\tilde{\omega}\big)\big]&=&\mathbb{E}_{\mathbb{P}\times \tilde{\mathbb{P}}}\big[\mathbb{E}_{\mathbb{P}\times \tilde{\mathbb{P}}}[U\big(X, Y,\tilde{\omega}\big)|\tilde{\mathcal{F}}]\big]=\mathbb{E}_{\mathbb{P}\times \tilde{\mathbb{P}}}\big[\mathbb{E}_{\mathbb{P}}[U\big(X, Y,\tilde{\omega}\big)]\big]=\mathbb{E}_{\mathbb{P}\times \tilde{\mathbb{P}}}\Big[\mathbb{E}_{\mathbb{P}}\big[\mathbb{E}_{\mathbb{P}}[U\big(X, Y,\tilde{\omega}\big)|Y]\big]\Big]\nonumber\\
		&\le&\mathbb{E}_{\mathbb{P}\times \tilde{\mathbb{P}}}\Big[\mathbb{E}_{\mathbb{P}}\big[U\big(\mathbb{E}_{\mathbb{P}}[X|Y], Y,\tilde{\omega}\big)\big]\Big]=\mathbb{E}_{\mathbb{P}\times \tilde{\mathbb{P}}}\Big[\mathbb{E}_{\mathbb{P}}\big[U\big(\mathbb{E}_{\mathbb{P}}[X], Y,\tilde{\omega}\big)\big]\Big]\nonumber\\
		&=&\mathbb{E}_{\mathbb{P}\times \tilde{\mathbb{P}}}\big[U\big(\mathbb{E}_{\mathbb{P}}[X], Y,\tilde{\omega}\big)\big],
	\end{eqnarray}
	where the first as well as the last equalities hold due to the tower rule for conditional expectations, the second equality follows from the statistical independence of $X,Y$ from $\tilde{\mathcal{F}}$, the inequality follows from the statewise concavity of $U(x,y,\omega)$ with respect to $x$ and Jensen's inequality, and the fourth equality holds due to the statistical independence between $X$ and $Y$.

\subsection{Proof of Corollary  \ref{cor1}}

	($i$) Let $\mathbb{P}(X=i,Y=j)=p_{ij}$, $i,j=0,1$. Since the correlation coefficient of $(X,Y)$ is non-negative, then $p_{00}p_{11}\ge p_{01}p_{10}$, which in turn implies
	$$\mathbb{E}_{\mathbb{P}}[X|Y=0]=\frac{p_{10}}{p_{00}+p_{10}}\le\frac{p_{11}}{p_{01}+p_{11}}=\mathbb{E}_{\mathbb{P}}[X|Y=1].$$ Thus, $\mathbb{E}_{\mathbb{P}}[X|Y=y]$ is non-decreasing with respect to $y$ so that $(X,Y)\in R^{\mathbb{P}}(X|Y)$. Then, it follows from Proposition \ref{cor:Uindependent-gamble-sufficient+}($i$) that $U$ is risk averse to $X$ in the presence of $Y$.
	
 ($ii$) In the case where $(X,Y)$ follow a bivariate normal distribution, $\mathbb{E}_{\mathbb{P}}[X|Y]=\ell Y+\eta$, where $\ell={\rm Cov}_{ \mathbb{P}}(X,Y)/$
 ${\rm Var}_{ \mathbb{P}}(Y)\ge 0$, where the inequality holds due to the non-negative correlation coefficient of $(X,Y)$. Thus, it follows from Proposition \ref{cor:Uindependent-gamble-sufficient+}($i$) that $U$ is risk averse to $X$ in the presence of $Y$.

\subsection{Proof of Theorem  \ref{thm:increasing-gamble-sufficient}}

	($i$) The sufficiency follows directly from Definition  \ref{uniform-risk-aversion-1} of uniform bivariate risk aversion and Proposition \ref{cor:Uindependent-gamble-sufficient++}.
	
	Now, we prove the necessity. 	Choose $Y(\omega)=c$ for all $\omega$ and for some constant $c\in\mathbb{R}$. Then for any risk  $X$, the random variables 
	$X$ and $Y$ are statistically independent.  Since for statistically  independent risks $X$ and $Y$, $U$ is uniformly risk averse to $X$ in the presence of  $Y$, then the inequality 
	$$\mathbb{E}_{\mathbb{P}\times \tilde{\mathbb{P}}}\big[U(X, c,\tilde{\omega})\big]\le   \mathbb{E}_{\tilde{\mathbb{P}}}\big[U(\mathbb{E}_{ \mathbb{P}}[X], c,\tilde{\omega})\big]$$
	holds for all $\mathbb{P}\times \tilde{\mathbb{P}}\in \mathcal{Q}$. 
	Letting $\omega_i$, $i=1,2$ be two arbitrary points in $\Omega$, $\tilde{\omega}_0$ be an arbitrary state in $\tilde{\Omega}$ and 
	$$\mathbb{P}\times\tilde{\mathbb{P}}=\lambda \delta_{\{\omega_1\times\tilde{\omega}_0\}}+(1-\lambda)\delta_{\{\omega_2\times\tilde{\omega}_0\}},$$
	where   $\lambda\in(0,1)$ and $\delta_{c}$ is the Dirac measure in the point $c$. The above inequality implies that for arbitrary $X(\omega_i)=x_i$, $i=1,2$,  
	$$\lambda U(x_1,c,\tilde{\omega}_0)+(1-\lambda) U(x_2,c,\tilde{\omega}_0)\le   U(\lambda x_1+(1-\lambda)x_2,c,\tilde{\omega}_0).$$
	Since $\tilde{\omega}_0$ is arbitrary, then the SDBUF $U(x,y,\tilde{\omega})$ is statewise concave in the first argument $x$. 
	
	($ii$) The sufficiency follows directly from the definition of uniform bivariate risk aversion  and Proposition \ref{cor:Uindependent-gamble-sufficient+}. 
	
	The necessity can be proved by contradiction from the following two steps. 
	
	\noindent ($ii$-1) Suppose that there exist $1\le M\le N$,  $\tilde{\omega}_0\in\tilde{\Omega}$ and   $(x_0,y_0)\in [\underline{x},\overline{x}]\times[\underline{y},\overline{y}]$ such that $(-1)^{1+M}U^{(1,M)}(x_0,y_0,\tilde{\omega}_0)> 0$. We only consider the case that $\underline{x}<x_0<\overline{x}$ and $\underline{y}<y_0<\overline{y}$,  since the other cases can be proved similarly.

Then, there exists $\epsilon_0>0$ such that for all $(x,y)\in[x_0-\epsilon_0, x_0+\epsilon_0]\times[y_0-\epsilon_0, y_0+\epsilon_0]$, $(-1)^{1+M}U^{(1,M)}(x,y,\tilde{\omega}_0)>0$. 
		Consider $Y,Z$ with $\mathbb{E}_{\mathbb{P}}\big[Z\big]=0$ such that $\mathbb{E}_{ \mathbb{P}}D_{k}(Z|\overline{y})=0$ for $k=2,\ldots,M-1$ and 
	\begin{equation}\label{eq:EDN}
		\mathbb{E}_{\mathbb{P}}D_{M}(Z|y)
		\begin{cases}
			>0, & y\in [y_0-\epsilon_0, y_0+\epsilon_0],\\
			=0, & {\rm otherwise}.
		\end{cases}
	\end{equation}	
	Then $(Z,Y)\in \mathbb{E}_{\mathbb{P}}D_{M}(Z|Y)$. From \eqref{dependence2}, we have that $(Z,Y)\in \mathbb{E}_{ \mathbb{P}}D_{N}(Z|Y)$ due to $M\le N$.
	
	Let $\hat x\in [x_0-\epsilon_0, x_0+\epsilon_0]$ and $\Delta>0$. For sufficient small $\Delta$, using the Taylor expansion of $\mathbb{E}_{ \mathbb{P}}[U(\hat x+\Delta\cdot Z,Y,\tilde{\omega}_0)]$ around $\hat x$, it yields that 
	\begin{equation}\label{eq:taylor}
		\mathbb{E}_{ \mathbb{P}}[U(\hat x+\Delta\cdot Z,Y,\tilde{\omega}_0)]=\mathbb{E}_{ \mathbb{P}}[U(\hat x,Y,\tilde{\omega}_0)]+\mathbb{E}_{ \mathbb{P}}[ZU^{(1,0)}(\hat x,Y,\tilde{\omega}_0)]\Delta+o(\Delta).
	\end{equation}
	Note that 
	\begin{eqnarray}
		&&\ \ \ \  \mathbb{E}_{\mathbb{P}}\big[ZU^{(1,0)}(\hat x,Y,\tilde{\omega}_0) \big]\nonumber \\
		&&=\mathbb{E}_{\mathbb{P}}[Z]\mathbb{E}_{\mathbb{P}}\big[U^{(1,0)}(\hat x,Y,\tilde{\omega}_0) \big]+{\rm Cov}_{\mathbb{P}}\big(Z, U^{(1,0)}(\hat x,Y,\tilde{\omega}_0) \big)\nonumber \\
		&&={\rm Cov}_{\mathbb{P}}\big(Z, U^{(1,0)}(\hat x,Y,\tilde{\omega}_0) \big)\nonumber \\
		&&=\sum_{k=2}^{M}(-1)^k U^{(1,k-1)}(\hat x,\overline{y},\tilde{\omega}_0) \mathbb{E}_{ \mathbb{P}}D_{k}(Z|\overline{y})+\int_{\underline{y}}^{\overline{y}} (-1)^{1+M}U^{(1,M)}(\hat x,y,\tilde{\omega}_0)\mathbb{E}_{ \mathbb{P}}D_{M}(Z|y) {\rm d} y, \label{eq:EPDN}\end{eqnarray}
	where the second equality holds due to that  $\mathbb{E}_{\mathbb{P}}[Z]=0$, and the third equality follows from \eqref{eq:sufficient-first2}. 
	Since $\mathbb{E}_{ \mathbb{P}}D_{k}(Z|\overline{y})=0$ for $k=2,\ldots,M-1$ and \eqref{eq:EDN} holds, then from the equality \eqref{eq:EPDN}, we know that 
	$$\mathbb{E}_{\mathbb{P}}\big[Z U^{(1,0)}(\hat x,Y,\tilde{\omega}_0) \big]=\int_{y_0-\epsilon_0}^{y_0+\epsilon_0} (-1)^{1+M}U^{(1,M)}(\hat x,y,\tilde{\omega}_0)\mathbb{E}_{ \mathbb{P}}D_{M}(Z|y) {\rm d} y>0,$$
	where the inequality follows from the assumption that $\mathbb{E}_{\mathbb{P}}D_{M}(Z|y)>0$ and $(-1)^{1+M}U^{(1,M)}(\hat x,y,\tilde{\omega}_0)>0$ for $y\in [y_0-\epsilon_0, y_0+\epsilon_0]$. Hence, combining the above inequality with   \eqref{eq:taylor},  we have that 
	\begin{equation}\label{ineq:E}
		\mathbb{E}_{\mathbb{P}}[U(\hat x+\Delta\cdot Z,Y,\tilde{\omega}_0)]>\mathbb{E}_{ \mathbb{P}}[U(\hat x,Y,\tilde{\omega}_0)].
	\end{equation}
	
	Define $X=\hat x+\Delta\cdot Z$. Since $(Z,Y)\in \mathbb{E}_{ \mathbb{P}}D_{N}(Z|Y)$, then $(X,Y)\in \mathbb{E}_{ \mathbb{P}}D_{N}(X|Y)$. 
	Letting  $\tilde{\mathbb{P}}=\delta_{\{\tilde{\omega}_0\}}$,  it  yields that 
	\begin{eqnarray}\label{eq:202293}
		&\mathbb{E}_{\mathbb{P}\times \tilde{\mathbb{P}}}\big[U(X, Y,\tilde{\omega})\big]&=\mathbb{E}_{ \mathbb{P}}\big[U(\hat x+\Delta\cdot Z, Y,\tilde{\omega}_0)\big]\nonumber\\
		&&> \mathbb{E}_{ \mathbb{P}}[U(\hat x,Y,\tilde{\omega}_0)]=\mathbb{E}_{ \mathbb{P}}[U(\mathbb{E}_{ \mathbb{P}}[X],Y,\tilde{\omega}_0)]= \mathbb{E}_{\mathbb{P}\times \tilde{\mathbb{P}}}\big[U(\mathbb{E}_{ \mathbb{P}}[X], Y,\tilde{\omega})\big],\end{eqnarray}
	where the inequality follows from \eqref{ineq:E} and the second equality holds due to that $\mathbb{E}_{ \mathbb{P}}[X]=\mathbb{E}_{ \mathbb{P}}[\hat{x}+\delta \cdot Z]=\hat x$. It leads to  a contradiction with the assumption that for any pair of risks $(X,Y)\in \mathbb{E}_{ \mathbb{P}}D_{N}(X|Y)$, $U$ is uniformly risk averse to $X$ in the presence of  $Y$.

	\noindent ($ii$-2) Suppose that there exist  $\tilde{\omega}_0\in\tilde{\Omega}$ and   $(x_0,y_0)\in [\underline{x},\overline{x}]\times[\underline{y},\overline{y}]$ such that $U^{(2,0)}(x_0,y_0,\tilde{\omega}_0)\ge  0$.  We still only consider the case that $\underline{x}<x_0<\overline{x}$ and $\underline{y}<y_0<\overline{y}$,  since the other cases can be proved by similar arguments. 
	
Then, there exists $\epsilon_0>0$ such that for all $(x,y)\in[x_0-\epsilon_0, x_0+\epsilon_0]\times[y_0-\epsilon_0, y_0+\epsilon_0]$, $U^{(2,0)}(x,y,\tilde{\omega}_0)\ge 0$. 
	
	Consider the random variable $Y$ with positive support on interval $[y_0-\epsilon_0, y_0+\epsilon_0]$  and the random variable $Z$ being statistically independent of $Y$ with $\mathbb{E}_{\mathbb{P}}\big[Z\big]=0$. 
	Then $(Z,Y)\in \mathbb{E}_{\mathbb{P}}D_{N}(Z|Y)$.
	
	Let $\hat x\in [x_0-\epsilon_0, x_0+\epsilon_0]$ and $\Delta>0$. For any sufficient small $\Delta$, using the Taylor expansion of $\mathbb{E}_{\mathbb{P}}[U(\hat x+\Delta\cdot Z,Y,\tilde{\omega}_0)]$ around $\hat x$, it yields that 
	\begin{eqnarray}
		&\mathbb{E}_{\mathbb{P}}[U(\hat x+\Delta\cdot Z,Y,\tilde{\omega}_0)]&=\mathbb{E}_{ \mathbb{P}}[U(\hat x,Y,\tilde{\omega}_0)]+\mathbb{E}_{ \mathbb{P}}[ZU^{(1,0)}(\hat x,Y,\tilde{\omega}_0)]\Delta+\frac{\mathbb{E}_{ \mathbb{P}}[U^{(2,0)}(\hat x,Y,\tilde{\omega}_0)Z^2]}{2}\Delta^2+o(\Delta^2)\nonumber \\
		&&=\mathbb{E}_{\mathbb{P}}[U(\hat x,Y,\tilde{\omega}_0)]+\frac{1}{2}\mathbb{E}_{ \mathbb{P}}[U^{(2,0)}(\hat x,Y,\tilde{\omega}_0)]\mathbb{E}_{ \mathbb{P}}[Z^2]\Delta^2+o(\Delta^2),\label{eq:taylor-2}
	\end{eqnarray}
	where the second equality follows from the assumptions that $Y,Z$ are statistically independent and  $\mathbb{E}_{\mathbb{P}}\big[Z\big]=0$. Since  $U^{(2,0)}(x,y,$
	$\tilde{\omega}_0)\ge 0$ 
	for any $(x,y)\in[x_0-\epsilon_0, x_0+\epsilon_0]\times[y_0-\epsilon_0, y_0+\epsilon_0]$, the random variable $Y$ has positive support on interval $[y_0-\epsilon_0, y_0+\epsilon_0]$, and $\hat x\in[x_0-\epsilon_0, x_0+\epsilon_0]$, then $\mathbb{E}_{ \mathbb{P}}[U^{(2,0)}(\hat x,Y,\tilde{\omega}_0)]\ge 0$. Combining this result with  \eqref{eq:taylor-2}, we have that $\mathbb{E}_{ \mathbb{P}}[U(\hat x+\Delta\cdot Z,Y,\tilde{\omega}_0)]\ge \mathbb{E}_{ \mathbb{P}}[U(\hat x,Y,\tilde{\omega}_0)]$.
	
	Define $X=\hat x+\Delta\cdot Z$. Since $(Z,Y)\in \mathbb{E}_{ \mathbb{P}}D_{N}(Z|Y)$, then $(X,Y)\in \mathbb{E}_{ \mathbb{P}}D_{N}(X|Y)$. Similar to the arguments in ($ii$-1), letting $\tilde{\mathbb{P}}=\delta_{\{\tilde{\omega}_0\}}$, we have that  for the pair of risks $(X,Y)$, the inequality $\mathbb{E}_{\mathbb{P}\times \tilde{\mathbb{P}}}\big[U(X, Y,\tilde{\omega})\big]$
	$\ge   \mathbb{E}_{\mathbb{P}\times \tilde{\mathbb{P}}}\big[U(\mathbb{E}_{ \mathbb{P}}[X], Y,\tilde{\omega})\big]$ holds, which leads to a contradiction with the assumption  that for any pair of risks $(X,Y)\in \mathbb{E}_{\mathbb{P}}D_{N}(X|Y)$, $U$ is uniformly risk averse to $X$ in the presence of $Y$. 
	
	($iii$) It can be proved by similar argument as that of the statement ($ii$), so we omit it.

	($iv$) It suffices to prove the necessity. Assume that for all possible pairs of risks $(X,Y)$, $U$ is uniformly  risk averse to  $X$ in the presence of $Y$.  
	
	Let $X$ and $Y$ be comonotonic random variables. Then  $(X,Y)\in \mathbb{E}_{\mathbb{P}}D_{1}(X|Y)$. From the assumption that $U$ is uniformly risk averse to $X$ in the presence of $Y$ and the statement ($ii$), we know that for each state  $\tilde{\omega}\in\tilde{\Omega}$ and for all $(x,y)\in [\underline{x},\overline{x}]\times[\underline{y},\overline{y}]$,  $U(x,y,\tilde{\omega})$ is statewise concave in its first argument $x$ and $U^{(1,1)}(x,y,\tilde{\omega})\le 0$. 
	
	Also, let $V(x,y,\tilde{\omega})=U(x,-y,\tilde{\omega})$ for each state  $\tilde{\omega}\in\tilde{\Omega}$ and for all $(x,y)\in [\underline{x},\overline{x}]\times[\underline{y},\overline{y}]$. Then, we have that 
	$$\mathbb{E}_{\mathbb{P}\times \tilde{\mathbb{P}}}\big[V(X, Y,\tilde{\omega})\big]=\mathbb{E}_{\mathbb{P}\times \tilde{\mathbb{P}}}\big[U(X, -Y,\tilde{\omega})\big]\le   \mathbb{E}_{\mathbb{P}\times \tilde{\mathbb{P}}}\big[U(\mathbb{E}_{ \mathbb{P}}[X], -Y,\tilde{\omega})\big]=\mathbb{E}_{\mathbb{P}\times \tilde{\mathbb{P}}}\big[V(\mathbb{E}_{ \mathbb{P}}[X], Y,\tilde{\omega})\big],$$
	where the inequality holds due to that 
	$U$ is uniformly bivariate risk averse to $(X,-Y)$. Then for $(X,Y)\in \mathbb{E}_{ \mathbb{P}}D_{1}(X|Y)$, $V$ is uniformly  risk averse to $X$ in the presence of $Y$, and thus from the statement ($ii$) in this theorem, we know that for each state  $\tilde{\omega}\in\tilde{\Omega}$ and for all $(x,y)\in [\underline{x},\overline{x}]\times[\underline{y},\overline{y}]$,  $V(x,y,\tilde{\omega})$ is concave in $x$ and $V^{(1,1)}(x,y,\tilde{\omega})\le  0$, which are  equivalent to the fact that $U(x,-y,\tilde{\omega})$ is statewise concave with respect to $x$ and $\frac{\partial^2   U(x,-y,\tilde{\omega})}{\partial x \partial y}\le 0$. Then, by Theorem 3 in \cite{1999On}, we have that for each fixed $\tilde{\omega}\in\tilde{\Omega}$ and for any  $x_1,x_2\in [\underline{x},\overline{x}]$, $y_1,y_2\in [\underline{y},\overline{y}]$,
	\begin{equation*}
		U(x_2,y_1,\tilde{\omega})-U(x_1,y_1,\tilde{\omega})=U(x_2,y_2,\tilde{\omega})-U(x_1,y_2,\tilde{\omega}).
	\end{equation*} 	
	Letting $x_1=0$, $x_2=x$, $y_1=y$ and $y_2=0$ in the above equation, and denoting $f(x,\tilde{\omega})=U(x,0,\tilde{\omega})-U(0,0,\tilde{\omega})$ and $g(y,\tilde{\omega})=U(0,y,\tilde{\omega})$, it yields that $U(x,y,\tilde{\omega})=f(x,\tilde{\omega})+g(y,\tilde{\omega})$. 
	Since $U(x,y,\tilde{\omega})$ is statewise concave in its first argument $x$, then  $f(x,\tilde{\omega})$ is statewise   concave in $x$. 

\subsection{Proof of Theorem  \ref{thm:independent-gamble-sufficient}}

From the assumption that for each state $\omega\in\Omega$ and for any  $(x,y)\in [\underline{x},\overline{x}]\times[\underline{y},\overline{y}]$, $U(x,y,\omega)$ is statewise concave with respect to $x$, we know that  for each state $\omega\in\Omega$, 
	$$U(X,Y,\omega)-U(\mathbb{E}_{\mathbb{P}}[X], Y,\omega)\le U^{(1,0)}(\mathbb{E}_{\mathbb{P}}[X], Y,\omega) (X-\mathbb{E}_{ \mathbb{P}}[X]). 
	$$
	Taking expectations on both side of the above inequality, it yields that
	\begin{equation}\label{eq:ex}
		\mathbb{E}_{\mathbb{P}}\big[U(X,Y,\omega) \big]-\mathbb{E}_{\mathbb{P}}\big[U(\mathbb{E}_{\mathbb{P}}[X],Y,\omega) \big]\le  \mathbb{E}_{\mathbb{P}}\big[U^{(1,0)}(\mathbb{E}_{\mathbb{P}}[X], Y,\omega) (X-\mathbb{E}_{\mathbb{P}}[X]) \big].
	\end{equation}
	Since $\mathbb{E}_{\mathbb{P}}\big[U^{(1,0)}(\mathbb{E}_{\mathbb{P}}[X], Y,\omega)]\mathbb{E}_{\mathbb{P}}\big[X-\mathbb{E}_{\mathbb{P}}[X]\big]=0$, then
	$$\mathbb{E}_{\mathbb{P}}\big[U^{(1,0)}(\mathbb{E}_{\mathbb{P}}[X], Y,\omega) (X-\mathbb{E}_{\mathbb{P}}[X]) \big] ={\rm Cov}_{\mathbb{P}}\big(X-\mathbb{E}_{\mathbb{P}}[X], U^{(1,0)}(\mathbb{E}_{\mathbb{P}}[X],Y,\omega)\big)={\rm Cov}_{\mathbb{P}}\big(X, U^{(1,0)}(\mathbb{E}_{\mathbb{P}}[X],Y,\omega)\big).$$
	Substituting the above equality into \eqref{eq:ex} and from the covariance condition that ${\rm Cov}_{\mathbb{P}}\big(X, U^{(1,0)}(\mathbb{E}_{\mathbb{P}}[X],Y,\omega)\big)\le 0$, 
	we have that $\mathbb{E}_{\mathbb{P}}\big[U(X,Y,\omega) \big]-\mathbb{E}_{\mathbb{P}}\big[U(\mathbb{E}_{\mathbb{P}}[X],Y,$ $\omega) \big]\le 0$. Hence, $U$ is risk averse to $X$ in the presence of $Y$. 

\subsection{Proof of Theorem  \ref{thm:expectation-dependent-gamble-sufficient-necessary}}
	
	\noindent ``(1) $\Rightarrow $ (2)''
	Since for each $\omega\in\Omega$, $U(x,y,\omega)$ is concave and  the covariance condition ${\rm Cov}_{\mathbb{P}}$
	$\big(X,U^{(1,0)}(\mathbb{E}_{\mathbb{P}}[X],Y,\omega)\big)\le 0$ holds for all $\mathbb{P}\in \mathcal{P}$, then for any pair of risks $(X,Y)\in \mathbb{E}_{\mathbb{P}}D_{N}(X|Y)$,  by Theorem  \ref{thm:independent-gamble-sufficient}, we know that the inequality $$\mathbb{E}_{\mathbb{P}}\big[U(X, Y,\omega)\big]\le \mathbb{E}_{\mathbb{P}}\big[U(\mathbb{E}_{\mathbb{P}}[X], Y,\omega)\big]$$ 
	holds for  all $\mathbb{P}\in \mathcal{P}$. Thus, from Definition \ref{uniform-risk-aversion}, we know that for any pair of risks $(X,Y)\in \mathbb{E}_{\mathbb{P}}D_{N}(X|Y)$, $U$ is uniform risk averse to $X$ in the presence of $Y$. 
	
	\noindent 	``(2) $\Rightarrow $ (3)'' It can be proved by contradiction. Suppose that the bivariate utility function $U(x,y,\omega)$ is state-dependent with the state space $\Omega$. 
	By the non-uniqueness of utility functions, for each  $\omega\in\Omega$, the utility function $U$ can be normalized as $U(\underline{x},\underline{y},\omega)=0$. Then, from the assumption that $x$ is a utility dominant attribute, we have that $U(\underline{x},y,\omega)=U(\underline{x},\underline{y},\omega)=0$ for each $\omega\in\Omega$ and for all $y\in[\underline{y},\overline{y}]$.

	Since $U(x,y,\omega)$ is state-dependent with $\Omega$, then there exist $\omega_1\neq  \omega_2\in\Omega$ such that $U(x_1,y_1,$
	$\omega_1)\neq U(x_1,y_1, \omega_2)$ for some $\underline{x}\le x_1\le \overline{x}$ and  $\underline{y}\le y_1\le \overline{y}$. Without loss of generality, assume that $U(x_1,y_1,\omega_1)< U(x_1,y_1, \omega_2)$. Then, there exists $\hat{x}\in(\underline{x}, x_1)$ such that   
	$
	U^{(1,0)}(\hat{x},y_1,\omega_1)<U^{(1,0)}(\hat{x},y_1,\omega_2)$ because if $U^{(1,0)}(x,y_1,\omega_1)\ge U^{(1,0)}(x,y_1,\omega_2)$ for all $x\in(\underline x,x_1)$, we	obtain a contradiction from
	$$U(x_1,y_1,\omega_1)=\int_{\underline{x}}^{x_1}U^{(1,0)}(x,y_1,\omega_1){\rm d}x\ge \int_{\underline{x}}^{x_1}U^{(1,0)}(x,y_1,\omega_2){\rm d}x=U(x_1,y_1,\omega_2),$$	
	where the equalities hold due to that $x$ is a utility dominant attribute such that $U(\underline{x},y_1,\omega_1)= U(\underline{x},y_1, \omega_2)=0$.
	
Hence, we know that there exists a closed ball $A(\hat{x},\epsilon)=\{(a,b)|(a-\hat{x})^2+(b-\hat{x})^2\le\epsilon \}$ such that for all $(a,b)\in A(\hat{x},\epsilon)$,
	\begin{equation}\label{eq:inequality}
		U^{(1,0)}(a,y_1,\omega_1)<U^{(1,0)}(b,y_1,\omega_2).
	\end{equation}
	
	Denote $$m(A,y_1)=\min\limits_{(a,b)\in A(\hat{x},\epsilon)}\{U^{(1,0)}(b,y_1,\omega_2)-U^{(1,0)}(a,y_1,\omega_1)\}.$$
	Let $\eta>0$ be a constant such that $[\hat{x}-\eta, \hat{x}+\eta] \in A(\hat{x},\epsilon)$. For each $\omega\in\Omega$, since $U^{(2,0)}(x,y,\omega)$ and $U^{(1,1)}(x,y,\omega)$ exist for all  $(x,y)\in[\underline{x},\overline{x}]\times[\underline{y},\overline{y}]$, then
	$U^{(1,0)}(x,y,\omega)$ is  uniformly continuous on  $[\underline{x},\overline{x}]\times[\underline{y},\overline{y}]$. Thus,    
	there exists  $\kappa>0$ such that for any $(x,y)\in [\hat{x}, \hat{x}+\eta]\times [y_1,y_1+\kappa]$,  
	\begin{equation}\label{ineq:gurantee}
		\big|U^{(1,0)}(x,y,\omega_2)-U^{(1,0)}(x,y_1,\omega_2)\big|< m(A,y_1),
	\end{equation}
	where $\kappa$ depends only on the value of $m(A,y_1)$ and is independent of  $(x,y)$. 
	
	Consider the probability space $(\Omega, \mathcal{F}, \mathbb{P})$, where $\Omega=\{\omega_1,\omega_2\}$ and $\mathbb{P}$ is the uniform distribution on $\{\omega_1,\omega_2\}$. Take $X(\omega_1)=\hat{x}-\eta$, $X(\omega_2)=\hat{x}+\eta$,  $Y(\omega_1)=y_1$ and $Y(\omega_2)=y_2$, where $y_1<y_2\le y_1+\kappa$. Then  $\mathbb{E}_{\mathbb{P}}[X]=\hat{x}$  and  $\mathbb{P}\big((X,Y)=(\hat{x}-\eta,y_1)\big)$=1/2=$\mathbb{P}\big((X,Y)=(\hat{x}+\eta,y_2)\big)$ such that $\mathbb{E}[X|Y=y_1]=\hat{x}-\eta<\hat{x}+\eta=\mathbb{E}[X|Y=y_2]$. Thus $(X,Y)\in R^{\mathbb{P}}(X|Y)$. From \eqref{dependence1} and \eqref{dependence2}, we know that
	$(X,Y)\in\mathbb{E}_{\mathbb{P}}D_{N}(X|Y)$. 
	
	Under the above conditions, we have that 
	$$ \mathbb{E}_{\mathbb{P}}\big[U\big(\mathbb{E}_{\mathbb{P}}[X(\omega)], Y(\omega),\omega\big)\big]=\frac{1}{2}U(\hat{x},y_1,\omega_1)+\frac{1}{2}U(\hat{x},y_2,\omega_2)$$
	and 
	$$
	\mathbb{E}_{\mathbb{P}}\big[U\big(X(\omega), Y(\omega),\omega\big)\big]=\frac{1}{2}U(\hat{x}-\eta,y_1,\omega_1)+\frac{1}{2}U(\hat{x}+\eta,y_2,\omega_2).$$
	Then
	\begin{eqnarray}\label{ineq:compare}
		&&\ \ \ \  \mathbb{E}_{\mathbb{P}}\big[U\big(X(\omega), Y(\omega),\omega\big)\big]- \mathbb{E}_{\mathbb{P}}\big[U\big(\mathbb{E}_{\mathbb{P}}[X(\omega)], Y(\omega),\omega\big)\big]\nonumber\\
		&&=\frac{1}{2}\big(U(\hat{x}+\eta,y_2,\omega_2)-U(\hat{x},y_2,\omega_2)\big)+\frac{1}{2}\big(U(\hat{x}-\eta,y_1,\omega_1)-U(\hat{x},y_1,\omega_1)\big)\nonumber\\
		&&=\frac{1}{2}\eta\big(U^{(1,0)}(\hat x_2,y_2,\omega_2)-U^{(1,0)}(\hat x_1,y_1,\omega_1)\big)\nonumber\\
		&&=\frac{1}{2}\eta\big(U^{(1,0)}(\hat x_2,y_1,\omega_2)-U^{(1,0)}(\hat  x_1,y_1,\omega_1)\big)+\frac{1}{2}\eta\big(U^{(1,0)}(\hat x_2,y_2,\omega_2)-U^{(1,0)}(\hat x_2,y_1,\omega_2)\big)		
		,\end{eqnarray}
	where $\hat x-\eta\le \hat x_1\le \hat{x}\le \hat x_2\le \hat{x}+\eta$ such that 
	$(\hat x_1, \hat x_2)\in A(\hat{x},\epsilon) $. By the inequality \eqref{eq:inequality} and from the definition of $m(A,y_1)$, we know that 
	\begin{equation}\label{eq:first}
		U^{(1,0)}(\hat x_2,y_1,\omega_2)-U^{(1,0)}(\hat x_1,y_1,\omega_1)\ge m(A,y_1)>0.
	\end{equation}
	Since $\hat x\le\hat x_2 \le \hat x+\eta$	and $y_1<y_2\le y_1+\kappa$, then from \eqref{ineq:gurantee}, we know that
	\begin{equation}\label{ineq:gurantee+}
		\big|U^{(1,0)}(\hat x_2,y_2,\omega_2)-U^{(1,0)}(\hat x_2,y_1,\omega_2)\big|< m(A,y_1).
	\end{equation}
%
Combining 	\eqref{ineq:compare}, \eqref{eq:first} and \eqref{ineq:gurantee+}, it yields that
\begin{eqnarray*}
	&&\ \ \ \  \mathbb{E}_{\mathbb{P}}\big[U\big(X(\omega), Y(\omega),\omega\big)\big]- \mathbb{E}_{\mathbb{P}}\big[U\big(\mathbb{E}_{\mathbb{P}}[X(\omega)], Y(\omega),\omega\big)\big]> 0.\end{eqnarray*}
Thus, for the considered distribution   $\mathbb{P}$ and  risks $(X,Y)\in \mathbb{E}_{\mathbb{P}}D_{N}(X|Y)$, the SDBUF  $U$ is not uniformly risk averse to $X$ in the presence of  $Y$. This is the desired contradiction.

\noindent ``(3) $\Rightarrow $ (1)'' If the bivariate utility function $U$ is state-independent, then $U(x,y,\omega)=U(x,y)$ for all $\omega\in\Omega$. Thus 
$${\rm Cov}_{\mathbb{P}}\big(X,U^{(1,0)}(\mathbb{E}_{\mathbb{P}}[X],Y,\omega)\big)={\rm Cov}_{\mathbb{P}}\big(X,U^{(1,0)}(\mathbb{E}_{\mathbb{P}}[X],Y)\big).
$$	
For all $\mathbb{P}\in \mathcal{P}$ and any pair of risks $(X,Y)\in \mathbb{E}_{\mathbb{P}}D_{N}(X|Y)$, from the conditions that for each $\omega\in\Omega$,   $U(x,y,\omega)$ is concave in $x$ and 
obeys risk apportionment of order $(1,k)$ for $k\in\{1,\ldots, N\}$,  
and by similar arguments as  \eqref{eq:sufficient-first2}, we have that $${\rm Cov}_{\mathbb{P}}\big(X,U^{(1,0)}(\mathbb{E}_{\mathbb{P}}[X],Y,\omega)\big)={\rm Cov}_{\mathbb{P}}\big(X,U^{(1,0)}(\mathbb{E}_{\mathbb{P}}[X],Y)\big)\le 0.$$

    \subsection*{Acknowledgement}

The authors acknowledge the support from the National Natural Science Foundation of China (Grants
No. 72271113) and the Natural Science Foundation of Jiangxi Province (Grants No. 20232ACB201003, 20232BAB201022). All authors share first authorship. 

\subsection*{Declarations}

\noindent {\bf Availability of Data} Data sharing is not applicable to this article as no new data were created or analyzed in this study.

\noindent {\bf Conflict of Interest} The authors declare no conflicts of interest nor competing interests.

\noindent {\bf Ethics Approval} Not applicable.

\noindent {\bf Informed Consent } Not applicable.

\newcommand{\DOI}[1]{Doi: \href{https://doi.org/#1}{#1}}
\bibliographystyle{elsarticle-harv}
\bibliography{compositive}

\begin{thebibliography}{46}
\expandafter\ifx\csname natexlab\endcsname\relax\def\natexlab#1{#1}\fi
\expandafter\ifx\csname url\endcsname\relax
  \def\url#1{\texttt{#1}}\fi
\expandafter\ifx\csname urlprefix\endcsname\relax\def\urlprefix{URL }\fi

\bibitem[{Abbas and Howard(2005)}]{2005Attribute}
Abbas, A.~E., Howard, R.~A., 2005. Attribute dominance utility. Decision
  Analysis 2, 185--206.

\bibitem[{Arrow(1971)}]{Essays1971}
Arrow, K.~J., 1971. Essays in the Theory of Risk-Bearing. Markham Publishing
  Company, Chicago.

\bibitem[{Baillon and Bleichrodt(2015)}]{2015Testing}
Baillon, A., Bleichrodt, H., 2015. Testing ambiguity models through the
  measurement of probabilities for gains and losses. American Economic Journal:
  Microeconomics 7, 77--100.

\bibitem[{Basak(2000)}]{2000A}
Basak, S., 2000. A model of dynamic equilibrium asset pricing with
  heterogeneous beliefs and extraneous risk. Journal of Economic Dynamics and
  Control 24, 63--95.

\bibitem[{Basak and Makarov(2014)}]{Basak2014Strategic}
Basak, S., Makarov, D., 2014. Strategic asset allocation in money management.
  The Journal of Finance 69~(1), 179--217.

\bibitem[{Chau et~al.(2017)Chau, Cosso, Fontana, and Mostovyi}]{2017Optimal}
Chau, H., Cosso, A., Fontana, C., Mostovyi, O., 2017. Optimal investment with
  intermediate consumption under no unbounded profit with bounded risk. Journal
  of Applied Probability 54, 710--719.

\bibitem[{Chiu(1996)}]{1996Risk}
Chiu, W.~H., 1996. Risk aversion with state-dependent preferences in the
  rank-dependent expected utility theory. The Geneva Papers on Risk and
  Insurance Theory 21, 159--177.

\bibitem[{Cook and Graham(1977)}]{1977The}
Cook, P.~J., Graham, D.~A., 1977. The demand for insurance and protection:
  {T}he case of irreplaceable commodities. The Quarterly Journal of Economics
  91, 143--156.

\bibitem[{Cuadras(2002)}]{2002On}
Cuadras, C.~M., 2002. On the covariance between functions. Journal of
  Multivariate Analysis 81, 19--27.

\bibitem[{Denuit et~al.(2011)Denuit, Eeckhoudt, and Menegatti}]{2011Correlated}
Denuit, M.~M., Eeckhoudt, L., Menegatti, M., 2011. Correlated riks, bivariate
  utility and optimal choices. Economic Theory 46, 39--54.

\bibitem[{Denuit and Mesfioui(2017)}]{2017Preserving}
Denuit, M.~M., Mesfioui, M., 2017. Preserving the rothschild-stiglitz type
  increase in risk with background risk: A characterization. Insurance:
  Mathematics and Economics 72, 1--5.

\bibitem[{Diamond and Stiglitz(1974)}]{1974Increases}
Diamond, P.~A., Stiglitz, J.~E., 1974. Increases in risk and in risk aversion.
  Journal of Economic Theory 8, 337--360.

\bibitem[{Eeckhoudt et~al.(2007)Eeckhoudt, Rey, and Schlesinger}]{2007A}
Eeckhoudt, L., Rey, B., Schlesinger, H., 2007. A good sign for multivariate
  risk taking. Management Science 53, 117--124.

\bibitem[{Eeckhoudt and Schlesinger(2006)}]{2006Putting}
Eeckhoudt, L., Schlesinger, H., 2006. Putting risk in its proper place.
  American Economic Review 96, 280--289.

\bibitem[{Finkelshtain et~al.(1999)Finkelshtain, Kella, and Scarsini}]{1999On}
Finkelshtain, I., Kella, O., Scarsini, M., 1999. On risk aversion with two
  risks. Journal of Mathematical Economics 31, 239--250.

\bibitem[{Franke et~al.(2006)Franke, Schlesinger, and
  Stapleton}]{2006Multiplicative}
Franke, G., Schlesinger, H., Stapleton, R.~C., 2006. Multiplicative background
  risk. Management Science 52, 146--153.

\bibitem[{Franke et~al.(2011)Franke, Schlesinger, and Stapleton}]{2011Risk}
Franke, G., Schlesinger, H., Stapleton, R.~C., 2011. Risk taking with additive
  and multiplicative background risks. Journal of Economic Theory 146,
  1547--1568.

\bibitem[{Guo et~al.(2016)Guo, Li, Liu, and Wang}]{2016Preserving}
Guo, X., Li, J., Liu, D., Wang, J., 2016. Preserving the rothschild-stiglitz
  type of increasing risk with background risk. Insurance: Mathematics and
  Economics 70, 144--149.

\bibitem[{Hill(2019)}]{2019A}
Hill, B., 2019. A non-{B}ayesian theory of state-dependent utility.
  Econometrica 87, 1341--1366.

\bibitem[{Howard(1980)}]{On1980}
Howard, R.~A., 1980. On Making Life and Death Decision. R. A. Howard, J. E.
  Matheson, eds. The Principles and Applications of Decision Analysis, Vol. II.
  Strategic Decisions Group, Menlo Park, CA.

\bibitem[{Hugonnier(2012)}]{2012Rational}
Hugonnier, J., 2012. Rational asset pricing bubbles and portfolio constraint.
  Journal of Economic Theory 147, 2260--2302.

\bibitem[{Ingersoll(1987)}]{Ingersoll1987}
Ingersoll, J., 1987. Theory of Financial Decision Making. Rowman \&
  Littlefield, Maryland.

\bibitem[{Jarrow(2017)}]{2017On}
Jarrow, R., 2017. On the existence of competitive equilibrium in frictionless
  and incomplete stochastic asset. Mathematics and Financial Economics 11,
  455--477.

\bibitem[{Jarrow and Li(2021)}]{2021Jarrow}
Jarrow, R., Li, S., 2021. Concavity, stochastic utility, and risk aversion.
  Finance and Stochastics 25, 311--330.

\bibitem[{Jokung(2011)}]{2011J}
Jokung, O., 2011. Risk apportionment via bivariate stochastic dominance.
  Journal of Mathematical Economics 47, 448--452.

\bibitem[{Karni(1983)}]{1983aversion}
Karni, E., 1983. Risk aversion for state-dependent utility functions:
  Measurement and applications. International Economic Review 24, 637--647.

\bibitem[{Karni et~al.(1983)Karni, Schmeidler, and Vind}]{1983On}
Karni, E., Schmeidler, D., Vind, K., 1983. On state dependent preferences and
  subjective probabilities. Econometrica 51, 1021--1031.

\bibitem[{Keeney and Raiffa(1976)}]{Keeney1976}
Keeney, R.~L., Raiffa, H., 1976. Decisions with multiple objectives:
  preferences and value trade-offs. Wiley, New York.

\bibitem[{Kelsey(1992)}]{1991Risk}
Kelsey, D., 1992. Risk and risk aversion for state-dependent utility. Theory
  and Decision 33, 71--82.

\bibitem[{Kelsey and Nordquist(1991)}]{1991A}
Kelsey, D., Nordquist, G.~L., 1991. A more general measure of risk aversion
  when utility is state-dependent. Oxford Economic Papers 43, 59--74.

\bibitem[{Li(2011)}]{2011The}
Li, J., 2011. The demand for a risky asset in the presence of a background
  risk. Journal of Economic Theory 146, 372--391.

\bibitem[{Li et~al.(2016)Li, Liu, and Wang}]{2016Risk}
Li, J., Liu, D., Wang, J., 2016. Risk aversion with two risks: A theoretical
  extension. Journal of Mathematical Economics 63, 100--105.

\bibitem[{Lu(2019)}]{2019Bayesian}
Lu, J., 2019. Bayesian identification: {A} theory for state-dependent
  utilities. American Economic Review 109, 3192--3228.

\bibitem[{Mostovyi(2017)}]{2017Mo}
Mostovyi, O., 2017. Optimal investment with intermediate consumption and random
  endowment. Mathematical Finance 27, 96--114.

\bibitem[{Nau(2003)}]{2003A}
Nau, R., 2003. A generalization of pratt-arrow measure to nonexpected-utility
  preferences and inseparable probability and utility. Management Science 49,
  1089--1104.

\bibitem[{Pratt(1964)}]{1964Pratt}
Pratt, J.~W., 1964. Risk aversion in the small and in the large. Econometrica
  32, 122--136.

\bibitem[{Pratt(1988)}]{1988Aversion}
Pratt, J.~W., 1988. Aversion to one risk in the presence of others. Journal of
  Risk and Uncertainty 1, 395--413.

\bibitem[{Ross(1981)}]{1981Some}
Ross, S., 1981. Some stronger measures of risk aversion in the small and in the
  large with applications. Econometrica 49, 621--638.

\bibitem[{Tsakas(2025)}]{2025Belief}
Tsakas, E., 2025. Belief identification by proxy. The Review of Economic
  Studies (forthcoming), https://doi.org/10.1093/restud/rdaf025.

\bibitem[{Tsetlin and Winkler(2005)}]{2005Risky}
Tsetlin, I., Winkler, R.~L., 2005. Risky choices and correlated background
  risk. Management Science 51, 1336--1345.

\bibitem[{\v{Z}itkovi\'{c}(2005)}]{Zitkovic2012Utility}
\v{Z}itkovi\'{c}, G., 2005. Utility maximization with a stochastic clock and an
  unbounded random endowment. Annals of Applied Probability 15, 748--777.

\bibitem[{Wong(2021)}]{2021Comparative}
Wong, K.~P., 2021. Comparative risk aversion with two risks. Journal of
  Mathematical Economics 97, 102536.

\bibitem[{Wright(1987)}]{1987Expectation}
Wright, R., 1987. Expectation dependence of random variables, with an
  application in portfolio theory. Theory and Decision 22, 111--124.

\bibitem[{Xie and Zhou(2022)}]{xie2022patchwork}
Xie, J.~H., Zhou, Z.~Y., 2022. Patchwork constructions of multiattribute
  utility functions. Decision Analysis 19~(2), 141--169.

\bibitem[{Yaari(1969)}]{1969Some}
Yaari, M., 1969. Some remarks on measures of risk aversion and on their uses.
  Journal of Economic Theory 1, 315--329.

\bibitem[{Zhu et~al.(2023)Zhu, Li, Yang, Xie, and Sun}]{zhu2023Asymptotic}
Zhu, W.~H., Li, L.~J., Yang, J.~P., Xie, J.~H., Sun, L.~L., 2023. Asymptotic
  subadditivity/superadditivity of value-at-risk under tail dependence.
  Mathematical Finance 33, 1314--1369.

\end{thebibliography}

\end{document}